# Deep learning-based noise reduction in low dose SPECT Myocardial Perfusion Imaging: Quantitative assessment and clinical performance

Narges Aghakhan Olia[1], Alireza Kamali-Asl[1], Sanaz Hariri Tabrizi[1], Parham Geramifar[2], Peyman Sheikhzadeh[3], Saeed Farzanefar[3] and Hossein Arabi[4]

[1] Department of Medical Radiation Engineering, Shahid Beheshti University, Tehran, Iran

[2] Research Center for Nuclear Medicine, Shariati Hospital, Tehran University of Medical Sciences, Tehran, Iran

[3] Department of Nuclear Medicine, Vali-Asr Hospital, Tehran University of Medical Sciences, Tehran, Iran

[4] Division of Nuclear Medicine and Molecular Imaging, Department of Medical Imaging, Geneva University Hospital, CH-1211 Geneva 4, Switzerland

# Abstract

**Purpose:** This work was set out to investigate the feasibility of SPECT-MPI dose reduction without sacrificing diagnostic accuracy. A deep learning approach was proposed to synthesize normal-dose images from the corresponding low-dose data at different reduced dose levels in the projection space.

**Methods:** Clinical SPECT-MPI images of 345 patients acquired from a dedicated cardiac SPECT in list-mode format were retrospectively employed to predict normal-dose images from low-dose data at the half, quarter, and one-eighth-dose levels. To simulate realistic low-dose projections, 50%, 25%, and 12.5% of the events were randomly selected from the list-mode data by applying a binomial subsampling. A generative adversarial network was implemented to predict non-gated normal-dose images in the projection space at the different reduced dose levels. Established metrics including the peak signal-to-noise ratio (PSNR), root mean squared error (RMSE), and structural similarity index metrics (SSIM) in addition to Pearson correlation coefficient analysis and derived parameters from Cedars-Sinai software were used to quantitatively assess the quality of the predicted normal-dose images. For clinical evaluation, the quality of the predicted normal-dose images was evaluated by a nuclear medicine specialist using a seven-point (-3 to +3) grading scheme.

**Results:** By considering PSNR, SSIM, and RMSE quantitative parameters among the different reduced dose levels, the highest PSNR ($42.49 \pm 2.37$) and SSIM ($0.99 \pm 0.01$), and the lowest RMSE ($1.99 \pm 0.63$) were obtained at the half-dose level in the reconstructed images. Pearson correlation coefficients were measured $0.997 \pm 0.001$, $0.994 \pm 0.003$, and $0.987 \pm 0.004$ for the predicted normal-dose images at the half, quarter, and one-eighth-dose levels, respectively. Regarding the normal-dose images as the reference, the Bland-Altman plots sketched for the Cedars-Sinai selected parameters exhibited remarkably less bias and variance in the predicted normal-dose images compared with the low-dose data at the entire reduced dose levels. Overall, considering the clinical assessment performed by a nuclear medicine specialist, 100%, 80%, and 11% of the predicted normal-dose images were clinically acceptable at the half, quarter, and one-eighth-dose levels, respectively.

**Conclusion:** Considering the quantitative metrics as well as the clinical assessment, the noise was effectively suppressed by the proposed network and the predicted normal-dose images were comparable to the reference normal-dose images at the half and quarter-dose levels. However, recovery of the underlying signals/information in low dose images beyond a quarter of the normal dose would not be feasible (due to very poor signal-to-noise-ratio) which will adversely affect the clinical interpretation of the resulting images.

# 1. Introduction

Single-photon emission computed tomography (SPECT) is a widespread nuclear medicine technique used in various imaging fields, including cardiac imaging [1, 2]. SPECT myocardial perfusion imaging (MPI) is an effective non-invasive method for diagnosis of coronary artery diseases, predicting disease progression, and evaluating acute coronary artery syndromes [3, 4]. To achieve high-quality images in nuclear medicine, a sufficient dose of radiopharmaceuticals needs to be injected, where reduced injected dose beyond the prescribed limit, would lead to poor signal-to-noise ratio (SNR), low-quality images, and limited diagnostic values [5, 6].

Since SPECT is considered the second-leading contributor to radiation dose among medical imaging modalities (with approximately 90% stress imaging annually performed in the United States), concerns about the radiation risk of this imaging modality have increased [7-10]. Multiple studies have been conducted to cope with the challenge of reducing the injected radiopharmaceuticals in nuclear medicine imaging without sacrificing the diagnostic/clinical value of the images. These are generally categorized into iterative image reconstruction, post-reconstruction filtering or post-processing, and machine learning methods [6, 11-13].

Iterative image reconstruction algorithms formulate low-dose image reconstruction as a convex optimization problem and suppress noise through statistical modeling of the signal formation and noise. Advanced iterative image reconstruction algorithms have shown that the injected dose or acquisition time could be decreased by a factor of two or higher in SPECT-MPI imaging [14-16]. In this regard, Ramon et al. quantified the accuracy of perfusion-defect detection in SPECT-MPI images as a function of injected dose to minimize administrated dose without sacrificing diagnostic performance [14]. The other approaches rely on the different post-processing and/or post-reconstruction denoising techniques such as nonlocal mean (NLM) or bilateral filters to suppress the noise in low-dose images [17, 18]. In addition to the aforementioned methods, which to some extent enable recovery of the underlying signals/structures in the low dose images, deep learning algorithms have exhibited promising performance/potentials in directly estimating/predicting high-quality standard-dose images from the corresponding low-dose ones [11, 19].

It has been shown that various types of deep neural networks are able to suppress the noise in low-dose computed tomography (CT) as well as positron emission tomography (PET) images leading to dependable estimation of the normal-dose images [6, 20-24]. Similarly, a number of studies have been conducted in the field of low-dose SPECT-MPI. In this regard, Ramon et al. demonstrated the feasibility of using several 3D convolutional denoising networks for SPECT-MPI denoising in the image domain at 1/2, 1/4, 1/8, and 1/16 of normal clinical dose levels [25]. Their proposed framework achieved comparable diagnostic accuracy in the predicted normal-dose images from the corresponding 1/2-dose images (compared to the reference full-dose images) for both conventional filtered back-projection (FBP) and ordered subset expectation maximization (OSEM) reconstruction methods. Song et al. investigated a 3D residual convolutional neural network (CNN) model to predict standard-dose images from 1/4-dose gated SPECT-MPI images [26]. Their results indicated that the proposed network effectively suppressed the noise levels in the myocardium and improved the spatial resolution of the left ventricular (LV) wall. Shiri et al. evaluated the potential of acquisition time reduction in SPECT-MPI using a residual network (ResNet) [27]. They followed two main approaches, reducing the number of projections and reducing the acquisition time per projection. The results demonstrated that the standard SPECT images estimated from the acquisition with reduced time per projection bear less quantification errors and more similarity to the reference images.

The aim of this study is to reduce the administered dose while preserving crucial/underlying structures without loss of diagnostic accuracy and clinical value in SPECT MPI images. Owing to the great success of deep neural networks in the field of normal-dose image estimation, we propose an end-to-end image translation approach to denoise the low-dose SPECT-MPI images. This work employs a deep generative adversarial network (GAN) model to estimate normal-dose images from the corresponding 1/2-, 1/4-, and 1/8-dose levels in an attempt to determine which reduced dose-level could be recovered by the GAN model with minimal loss of image quality and clinical value.

## 2. Materials & Methods

### 2.1. Data Acquisition

SPECT-MPI data were acquired from 345 patients (193 female and 152 male) scanned by the ProSPECT scanner (Parto Negar Persia, Iran), a dedicated cardiac SPECT with dual-head fixed 90° angle detectors. To prevent radiopharmaceutical re-injection, data acquisition was carried out in list-mode format to simulate the corresponding low-dose imaging. Using a two-day rest/stress acquisition protocol, image acquisition was conducted approximately one hour after the injection of $22 \pm 3$ mCi of technetium 99m-sestamibi. To reduce breast tissue and diaphragm attenuation, women and men underwent supine and prone imaging, respectively. The acquisition protocol consisted of 32 projections with 20 to 25 seconds per projection from right anterior oblique (RAO) to left posterior oblique (LPO). According to the electrocardiography (ECG) signal collected within the acquisition, detected photons were registered into 8 gate intervals during a cardiac cycle.

To simulate half-dose, quarter-dose, and one-eighth-dose acquisitions, regardless of the temporal information, the number of detected photons was reduced by applying a binomial subsampling. In this subsampling method, each registered photon in the projection space would be either kept or rejected through a probability function mimicking the different low dose levels.

The software dedicatedly developed for the ProSPECT scanner was employed to convert the list-mode data to non-gated projection data (64 × 64 × 32 voxels) and gated projection data (64 × 64 ×256 voxels) with a voxel size of 6.4 × 6.4 × 6.4 mm$^3$.

### 2.2. Data Preparation

Since the count rate from the liver absorption in SPECT-MPI is relatively high, projection images were manually cropped to exclude the liver from the cardiac images by a physician. The distinction between heart and liver was not possible for 15 patients, so we excluded them from the dataset. Projection data of 295 patients were randomly selected as the training and the remaining 35 patients were used as an external test dataset to assess the performance of the GAN model. According to the clinical report of the SPECT-MPI images, patients were divided into four groups: healthy, low-risk, intermediate-risk, and severe-risk. In this light, to fairly evaluate the network performance, the test dataset included 8, 16, 6, and 5 samples from these groups, respectively.

### 2.3. Deep Network Architecture

The GAN architecture is composed of a generator network to predict/estimate normal-dose images and a discriminator network that classifies the synthesized images as real or fake [28]. These networks are trained concurrently in an adversarial process to compete with each other. The discriminator weights are updated independently, while the generator model is updated via the discriminator feedback (Supplemental Figure 1).

#### 2.3.1. Generator Network

The generator network in this architecture is an encoder-decoder model (U-Net) shown in Figure 1. This model utilizes low-dose images as the input to estimate normal-dose images; it encodes the input image to the bottleneck layer, then decodes the data from the bottleneck layer to synthesize the output image. In this network, skip connections are used between the corresponding encoder and decoder layers.

In the encoding path, the input layer is followed by six encoder blocks. The number of 4 × 4 kernels with stride 2 in the encoder blocks are 64, 128, 256, 512, 512, and 512. In the second to fourth encoder blocks, the Batch

Normalization layer is used after the convolutional layers. These layers are followed by the Leaky ReLU (with slope 0.2) activation function in the first five encoder blocks. Likewise, ReLU activation function is used in the sixth encoder block.

After the bottleneck layer, six decoder blocks are used in the decoding path. In these blocks, according to the defined encoders, the number of feature maps decreases from 512 to 1. Each block consists of a 4 × 4 kernel in the deconvolution layer by a stride of 2 in each direction, followed by a Batch Normalization layer. In the first five decoders, skip connections are used to concatenate the data from each layer in the encoder path to the corresponding layer in the decoder path. These shortcut connections are aimed to prevent the gradient vanishing issue that may occur in complicated deep neural networks. Finally, concatenated results are passed through a ReLU activation function. In the last decoder, the defined deconvolutional layer is followed by the sigmoid activation function. Empirically, in the first decoder block, we use a drop-out layer to prevent overfitting. Due to the fact that the pooling layers reduce the spatial resolution of the input images, these layers were not considered in this architecture to avoid any feature/information loss throughout the synthesis process.

The generator is updated via a weighted sum of both the adversarial loss and the L2-norm loss. The update of the trainable parameters is carried out to minimize the L2- norm loss calculated between the predicted normal-dose and the reference normal-dose images. The L2-norm loss was selected as resulted in a high-quality synthesis of the standard-dose SPECT images. Besides, via using adversarial loss, the generator weights are updated to minimize the loss of the discriminator (to better distinguish between real or fake samples) leading to overall better performance of the GAN model to produce more realistic images. Within the training process, a weighting factor of 100/1 was optimized in favor of the L2-norm loss, leading to overall peak performance of the GAN model.

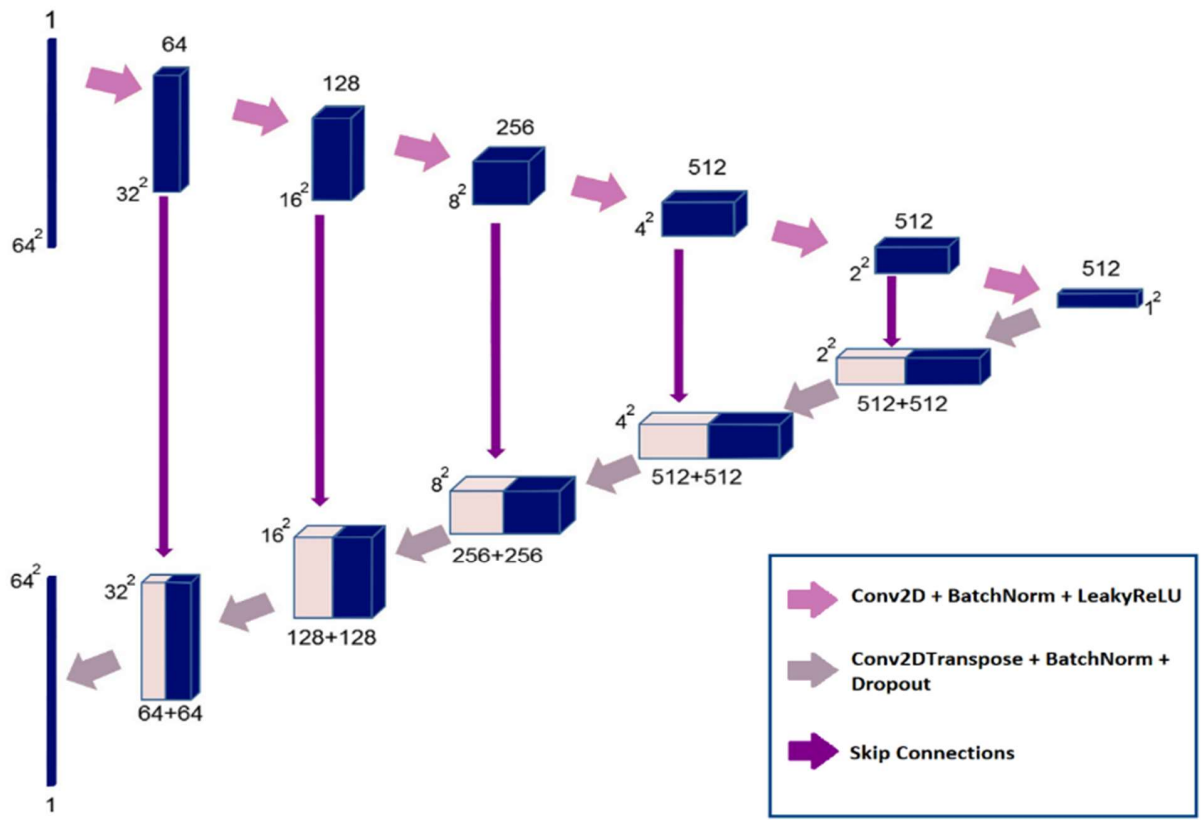

**Figure 1.** Architecture of the generator network in the GAN model.

### 2.3.2. Discriminator Network

The discriminator network, serving as an image classifier, takes both of the low-dose and the standard-dose images (either reference or synthesized normal-dose) as input to determine whether the normal-dose image is a real or a fake translation of the low-dose image. Figure 2 illustrates the architecture of the discriminator.

Regarding Figure 2, the network consists of a concatenate layer and five convolutional blocks. The number of 4 × 4 kernels with stride 2 applied in the first convolutional block is 48, and this number is doubled at each three following convolutional blocks, while the stride step in the fourth convolutional block becomes 1. The 2D convolutional layer is followed by the Batch Normalization layer and Leaky ReLU (with slope 0.2) activation function in each of the four convolutional blocks. Finally, the data is passed through a 1 × 1 single-filter convolutional layer, a Batch Normalization layer, and a sigmoid activation function. The Binary Cross-Entropy loss function was used for the training of the model with about 50 epochs.

The network was implemented using the Keras deep learning framework based on the TensorFlow libraries in Python 3.7. All the experiments were carried out on NVIDIA GeForce GTX 1060 with a 6GB memory graphical processing unit. Adaptive moment estimation (Adam) optimizer with a learning rate of 0.001 was used to minimize the loss functions.

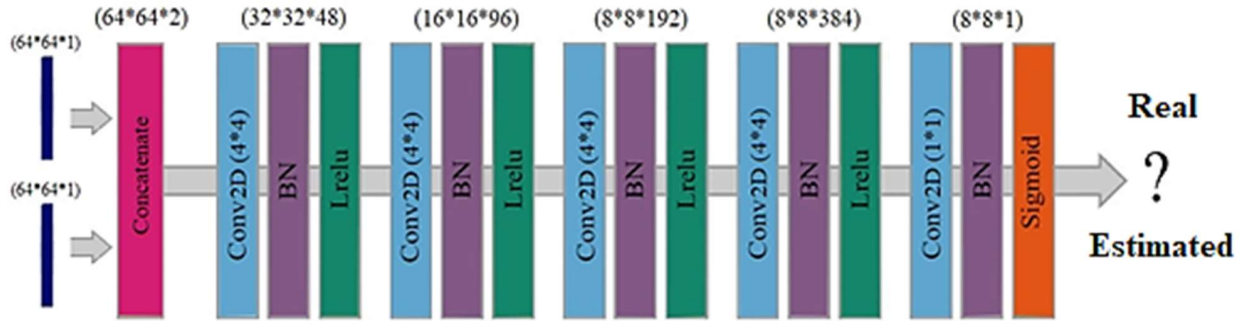

**Figure 2.** Architecture of the discriminator network in the GAN model. Conv2D: 2D convolutional layer, BN: Batch Normalization layer, Lrelu: Leaky ReLU activation function.

## 2.4. Image Reconstruction

The low-dose projection data (for 1/2, 1/4, and 1/8 levels) obtained from a random sampling of the list-mode acquisition, were employed for the training of the GAN model considering the standard-dose projection data as reference. The model was trained and evaluated separately for non-gated half-dose to normal-dose, non-gated quarter-dose to normal-dose, non-gated one-eighth-dose to normal-dose, and gated half-dose to normal-dose.

Normal-dose, low-dose, and predicted normal-dose projection data from the test dataset were reconstructed using OSEM algorithm by Cedars-Sinai software in three standard cardiac planes; short-axis (SA), vertical long-axis (VLA), and horizontal long-axis (HLA). The OSEM parameters were set to 8 iterations and 2 subsets. Furthermore, we applied a post-smoothing Butterworth filter with cutoff = 0.45 and order = 10.

## 2.5. Assessment Strategies
### 2.5.1. Quantitative Analysis

The quality of predicted normal-dose data, either in the projection or image space, was assessed using the standard quantitative metrics, including peak signal-to-noise ratio (PSNR), root mean squared error (RMSE), and structural similarity index metrics (SSIM) quantitative metrics are presented in equations 1-3, respectively, considering the normal-dose data as reference. Moreover, these metrics were also calculated for the low-dose images to provide a base-line for performance assessment of the GAN model.

$$\text{PSNR(dB)} = 20 \log_{10}\left(\frac{\text{Peak}}{\text{MSE}}\right) \tag{1}$$

$$\text{RMSE} = \sqrt{\frac{1}{n}\sum_{i=1}^{n}(y_i - \tilde{y}_i)^2} \tag{2}$$

$$\text{SSIM} = \frac{(2\mu_y\mu_{\tilde{y}}+C_1)(2\delta_{y,\tilde{y}}+C_2)}{(\mu_y^2+\mu_{\tilde{y}}^2+C_1)(\delta_y^2+\delta_{\tilde{y}}^2+C_2)} \tag{3}$$

In Eq. (1), Peak indicates the maximum count of either predicted normal-dose or low-dose data, and *MSE* stands for mean square error. In Eq. (2), *n* and *i* denote the total number of voxels and voxel index, respectively. *y* indicates the normal-dose data and $\tilde{y}$ is either the synthetic or low-dose data. $\mu_y$ and $\mu_{\tilde{y}}$ in Eq. (3) denote the mean values of the reference and synthetic/low-dose images, respectively. $\delta_{y,\tilde{y}}$ indicates the covariance of $\delta_y$ and $\delta_{\tilde{y}}$, which in turn represent the variances of the normal-dose and predicted normal-dose/low-dose images, respectively. The constant parameters $C_1$ and $C_2$ ($C_1$=0.01 $and$ $C_2$=0.02) were set to avoid division by very small values.

### 2.5.2. Cedars-Sinai Quantitative Parameters

Extent, Summed Stress Percent (SS%) or Summed Rest Percent (SR%), Summed Stress Score (SSS) or Summed Rest Score (SRS), Total Perfusion Deficit (TPD%), Volume, Wall, Shape Eccentricity, and Shape Index were calculated using Quantitative Perfusion SPECT (QPS) package in Cedars-Sinai software. The abovementioned metrics were calculated on the reconstructed reference, low-dose, and predicted normal-dose SPECT images using the standard reconstruction settings used in clinical routine. Afterward, Bland-Altman plots were sketched to describe the agreement between the predicted normal-dose/low-dose and reference normal-dose data. Finally, the Pearson correlation coefficient was computed for the derived parameters according to Eq. (4).

$$\rho = \frac{\sum_{i=1}^{n}(y_i-\mu_y)(\tilde{y}_i-\mu_{\tilde{y}})}{\sqrt{\sum_{i=1}^{n}(y_i-\mu_y)^2}\sqrt{\sum_{i=1}^{n}(\tilde{y}_i-\mu_{\tilde{y}})^2}} \tag{4}$$

### 2.5.3. Clinical Evaluation

The Summed Score (SS) parameter was calculated for the low-dose, predicted normal-dose, and reference normal-dose reconstructed images in the test dataset by a nuclear medicine specialist. Afterward, to express diagnostic differences in the predicted normal-dose/low-dose SPECT images with respect to the normal-dose ground truth, a scoring scheme with a range of between -3 to +3 was employed by a physician, wherein 0 is equivalent to no diagnostic changes, and ± 3 is equivalent to considerable changes compared to the reference normal-dose data. Positive numbers indicate higher activity estimation, and negative numbers indicate lower activity estimation compared to the reference normal-dose images. Finally, the Pearson correlation coefficient was calculated between the reference normal-dose and the predicted normal-dose/low-dose images.

## 3. Results
### 3.1. Qualitative Assessment

The predicted normal-dose images in both projection and image domains exhibited considerable improvement in image quality comparing to the low-dose images. Figure 4 depicts the predicted non-gated normal-dose projections for the different low-dose levels. The visual inspection reveals that at half-dose, compared to the quarter-dose and one-eighth-dose, the GAN model achieved nearly the same image quality as the reference normal-dose data. Image quality improvement is apparent for the predicted projections at the quarter-dose level, however, increased signal loss is observed in predicted projections from one-eighth-dose data. Figure 5 displays the SA, VLA, and HLA views of the reconstructed non-gated SPECT-MPI images, including reference normal-dose, low-dose, and predicted normal-dose for a representative subject with severe-risk diagnosis. It is shown that noise is appropriately suppressed at different reduced dose levels, where the LV wall appears more uniform/natural. Generally, the predicted images exhibited good agreement to the reference normal-dose data, whereas notable signal loss and or noise-induced pseudo signals are seen in the low-dose images. The reconstructed non-gated images for the patients diagnosed with normal perfusion, low-risk, and intermediate-risk are presented in supplemental Figure 2-4.

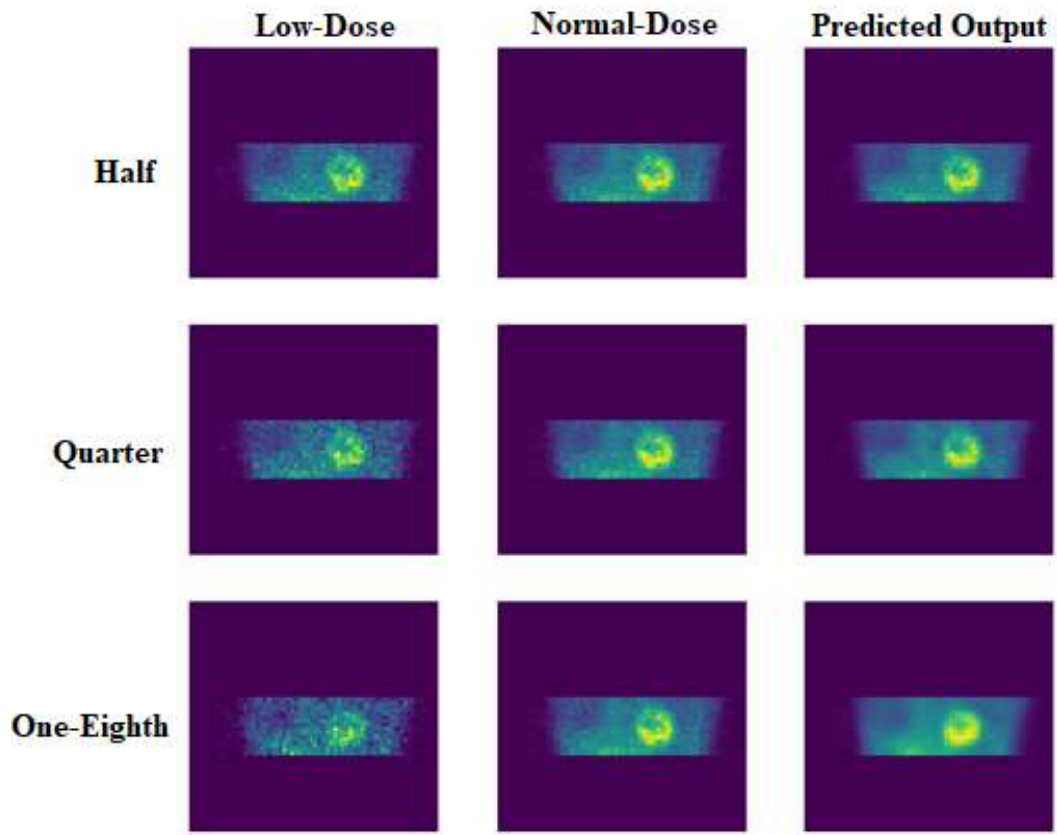

**Figure 4.** Illustration of the predicted non-gated projections for a randomly selected patient from the test dataset at the half-, quarter, and one-eighth-dose levels compared to the reference normal-dose and low-dose projections.

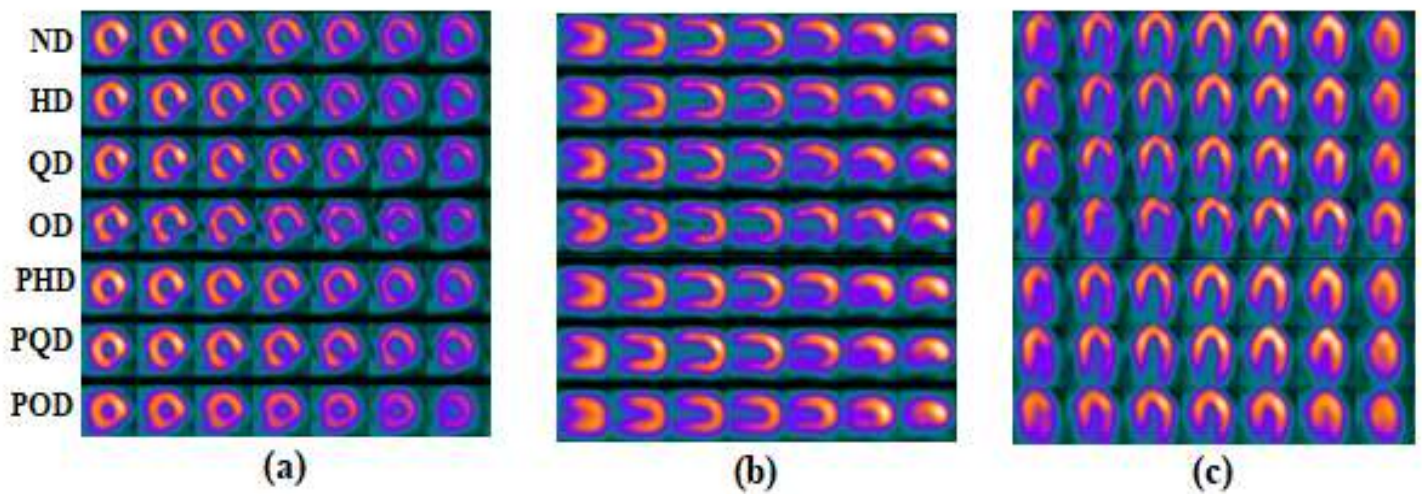

**Figure 5**. Reconstructed non-gated images for a patient with severe-risk. (a): short-axis view, (b): long vertical-axis view, (c): horizontal long-axis view. In (a), (b), and (c), the rows from top to bottom correspond to the normal-dose (ND), half-dose (HD), quarter-dose (QD), one-eighth-dose (OD), predicted half-dose (PHD), predicted quarter-dose (PQD), and predicted one-eighth-dose (POD), respectively.

## 3.2. Quantitative Analysis

Regarding the normal-dose images as the reference, PSNR, SSIM, and RMSE were calculated separately for the entire 35 patients in the test dataset for the non-gated predicted as well as low-dose data using Eq. (1) to Eq. (3). The mean and standard deviation of these metrics calculated in the projection and image spaces are reported in Tables 1 and 2, respectively. Additionally, the paired sample t-test at the 5% significance level was conducted between the PSNR, SSIM, and RMSE metrics obtained from the low-dose and the corresponding predicted normal-dose data. The *p*-values obtained from the statistical test are also presented in Tables 1 and 2.

Regarding the values obtained from the low-dose and predicted normal-dose projections in Table 1, there was a substantial increase in PSNR metric (12.4%, 25.2%, and 32.1%) for the predicted projections from half, quarter, and one-eighth dose levels, respectively. The SSIM increased by 2.1%, 4.3%, and 6.7%, whereas that RMSE decreased markedly by 38.5%, 56.3%, and 59.7% for the predicted projections from half, quarter, and one-eighth dose levels, respectively. Table 2 summarizes the quantitative analysis results of PSNR, RMSE, and SSIM metrics in the image space. The predicted images at the half-dose level achieved the highest SSIM (0.99 ± 0.01) and PSNR (42.49 ± 2.37), and the lowest RMSE (1.99 ± 0.63) with respect to the reference normal-dose images, while the predicted images at the one-eighth dose level resulted in the lowest PSNR (33.44 ± 2.63) and SSIM (0.95 ± 0.02), and the highest RMSE (5.70 ± 1.90) compared to the reference normal-dose images.

The null hypothesis of the t-test was rejected in the projection and image space for most of the cases with low *p*-values. The box plots of these quantitative metrics are presented in supplemental Figure 5. Furthermore, the image quality was quantified using the Pearson correlation coefficient calculated from Eq. (4) for the low-dose and predicted normal-dose reconstructed images versus the reference normal-dose counterparts. Figure 6 shows the mean and standard deviation of the Pearson correlation coefficients obtained from 35 patients in the non-gated test dataset for the entire reduced dose levels. Regarding Figure 6, the mean of Pearson correlation coefficient increased up to 1 for the predicted images as dose level increased from 1/8 to 1/2 wherein a significant decrease in standard deviation was observed for the entire dose levels compared to the corresponding low-dose images. For instance, the predicted normal-dose images yielded $\rho = 0.994 \pm 0.003$ compared to $\rho = 0.987 \pm 0.007$ obtained from the low dose images at the quarter-dose level.

**Table 1.** Quantitative results associated with the different dose levels in the projection space. *p*-value between the low-dose and the predicted normal-dose projections at each reduced dose level is presented.

| Parameters | Projection Space | | | | | |
|---|---|---|---|---|---|---|
| | HD | PHD | QD | PQD | OD | POD |
| PSNR | 33.22 ± 1.93 | 37.34 ± 1.33 | 28.25 ± 1.99 | 35.37 ± 1.56 | 24.51 ± 1.69 | 32.37 ± 1.77 |
| | *p*-value < 0.001 | | *p*-value < 0.001 | | *p*-value < 0.001 | |
| SSIM | 0.96 ± 0.01 | 0.98 ± 0.01 | 0.93 ± 0.01 | 0.97 ± 0.01 | 0.89 ± 0.02 | 0.95 ± 0.01 |
| | *p*-value < 0.001 | | *p*-value < 0.001 | | *p*-value < 0.001 | |
| RMSE | 5.69 ± 1.19 | 3.50 ± 0.52 | 10.10 ± 2.19 | 4.41 ± 0.78 | 15.52 ± 3.26 | 6.26 ± 1.23 |
| | *p*-value < 0.001 | | *p*-value < 0.001 | | *p*-value < 0.001 | |

HD: Half-Dose, PHD: Predicted Half-Dose, QD: Quarter-Dose, PQD: Predicted Quarter-Dose, OD: One-eighth-Dose, POD: Predicted One-eighth-Dose.

**Table 2.** Quantitative results associated with different dose levels in the image space. *p*-value between the low-dose and the predicted normal-dose projections at each reduced dose level.

| Parameters | Image Space | | | | | |
|---|---|---|---|---|---|---|
| | HD | PHD | QD | PQD | OD | POD |
| PSNR | 39.12 ± 2.68 | 42.49 ± 2.37 | 35.75 ± 2.72 | 40.17 ± 2.89 | 32.09 ± 2.74 | 33.44 ± 2.63 |
| | *p*-value < 0.001 | | *p*-value < 0.001 | | *p*-value < 0.04 | |
| SSIM | 0.97 ± 0.01 | 0.99 ± 0.01 | 0.95 ± 0.02 | 0.98 ± 0.01 | 0.90 ± 0.04 | 0.95 ± 0.02 |
| | *p*-value < 0.001 | | *p*-value < 0.001 | | *p*-value < 0.001 | |
| RMSE | 2.94 ± 0.96 | 1.99 ± 0.63 | 4.37 ± 1.48 | 2.93 ± 1.19 | 6.66 ± 2.04 | 5.70 ± 1.90 |
| | *p*-value < 0.001 | | *p*-value < 0.001 | | *p*-value < 0.05 | |

HD: Half-Dose, PHD: Predicted Half-Dose, QD: Quarter-Dose, PQD: Predicted Quarter-Dose, OD: One-eighth-Dose, POD: Predicted One-eighth-Dose.

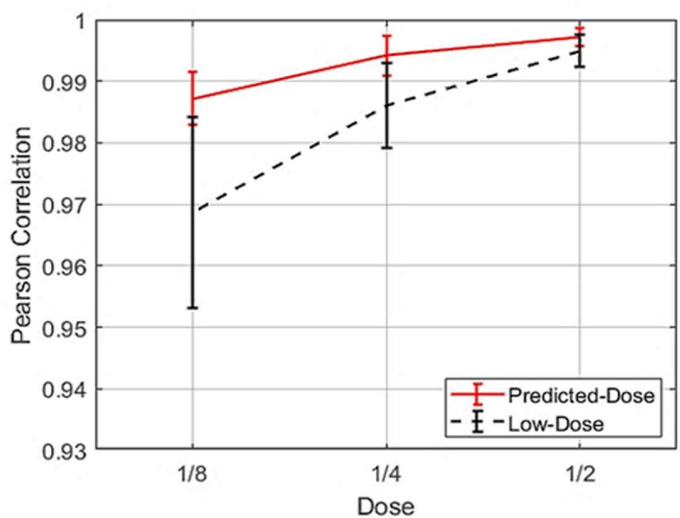

**Figure 6.** Comparing the Pearson correlation coefficients obtained from the low-dose and predicted-normal-dose reconstructed images at the half, quarter, and one-eighth dose levels.

### 3.3. Cedars-Sinai Quantitative Parameters

The quantitative accuracy of the non-gated synthetic images was further investigated using QPS software indices including Defect, Extent, SS% or SR%, SSS or SRS, TPD%, Volume, Wall, Shape Eccentricity, and Shape Index. The Bland-Altman plot was employed to present the derived indices from the low-dose and predicted standard-dose reconstructed images together with Pearson correlation analysis. The Bland-Altman plots of SSS/SRS and TPD% quantitative parameters are presented for the different dose levels in Figures 7 and 8. Considering the SSS/SRS index shown in Figure 7, the Bland-Altman plots displayed the lowest bias (0.17) and variance (95% CI: -1.02, +1.36) for the predicted half-dose images compared with the reference normal-dose data. At the quarter-dose level, the data points associated with the low-dose images exhibited variation nearly twice as large as predicted normal-dose data. At the one-eighth dose level, despite the large dispersion of the data point, closer agreement was observed between the predicted normal-dose and reference normal-dose data in comparison with the low-dose data. Likewise, regarding the TPD% index presented in Figure 8, less bias and variance were observed in the predicted normal-dose images than the low-dose data at the entire reduced dose levels. The Bland-Altman plots for the rest of the indices can be found in supplemental Figures 6-12. Moreover, the box plots of these indices at the different dose levels are shown in supplemental Figure 13.

Table 3 presents the Pearson correlation coefficients for the QPS indices. Despite the overall improvements seen for the predicted half-dose and quarter-dose images, no significant improvement was observed for the one-eighth-dose level due to the extremely high noise level.

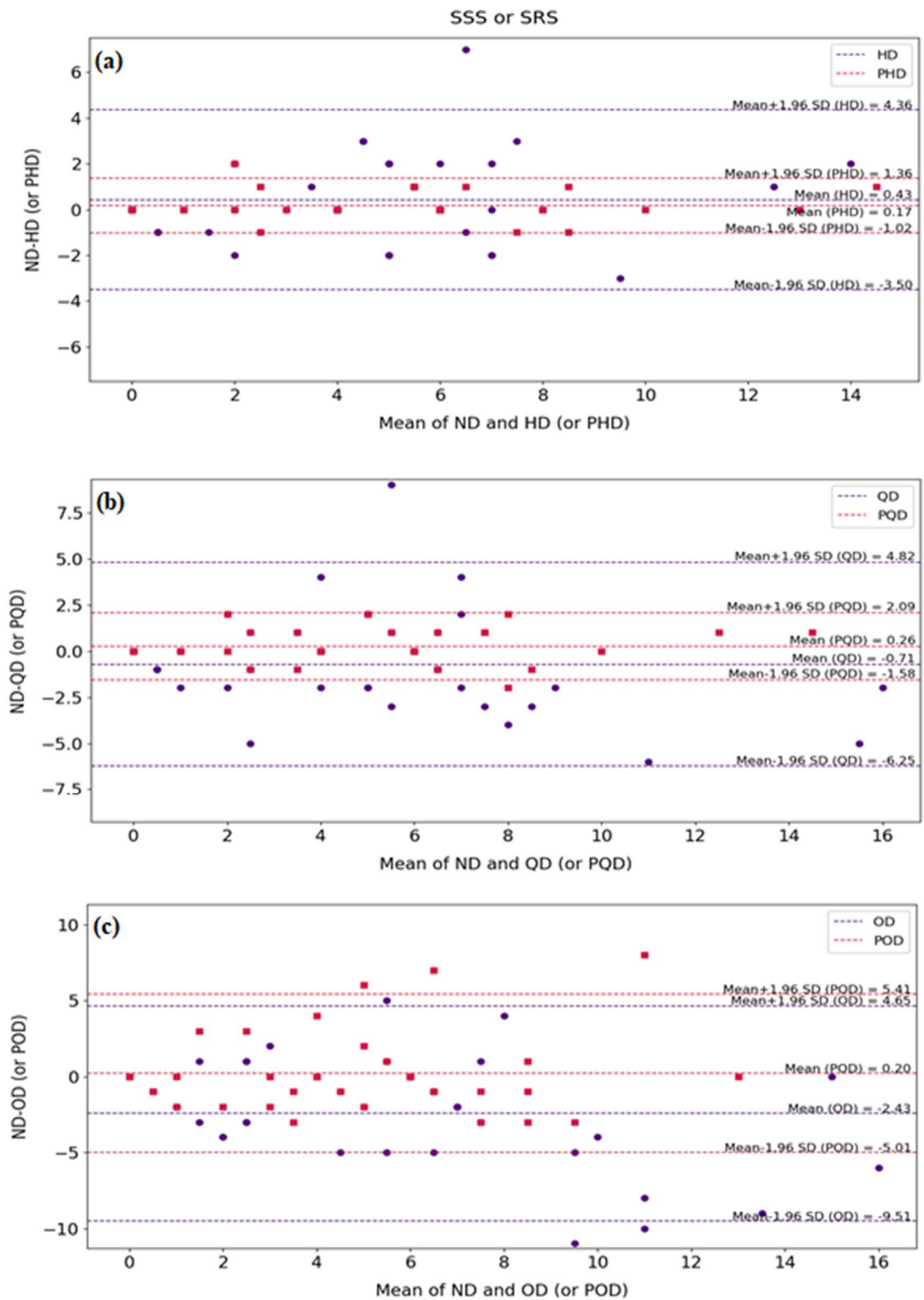

**Figure 7.** Bland-Altman plots of SSS index for the low-dose and predicted normal-dose images at (a) half-dose, (b) quarter-dose level, and (c) one-eighth-dose levels compared with the reference normal-dose images. The blue and red dashed lines indicate the mean and 95% confidence interval of the SSS differences in the low-dose and predicted normal-dose images, respectively. HD: Half-Dose, PHD: Predicted Half-Dose, QD: Quarter-Dose, PQD: Predicted Quarter-Dose, OD: One-eighth-Dose, POD: Predicted One-eighth-Dose.

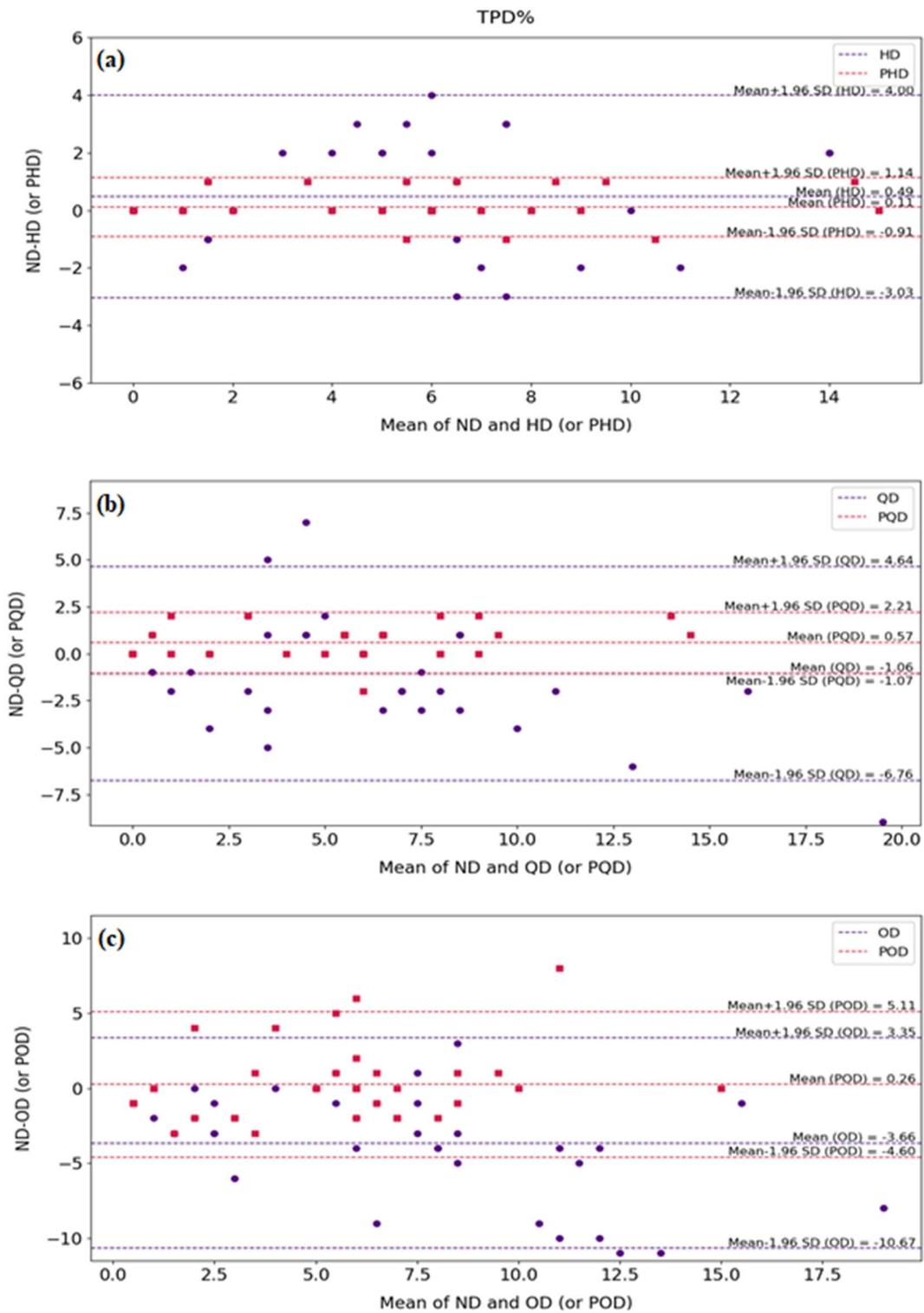

**Figure 8.** Bland-Altman plots of TPD% index for the low-dose and predicted normal-dose images at (a) half-dose, (b) quarter-dose, and (c) one-eighth-dose levels compared with the reference normal-dose images. The blue and red dashed lines designate the mean and 95% confidence interval of the TPD% differences in the low-dose and predicted normal-dose images, respectively. HD: Half-Dose, PHD: Predicted Half-Dose, QD: Quarter-Dose, PQD: Predicted Quarter-Dose, OD: One-eighth-Dose, POD: Predicted One-eighth-Dose.

**Table 3.** Pearson correlation coefficients of the QPS quantitative parameters considering the real normal-dose images as the reference.

| Parameters | HD | PHD | QD | PQD | OD | POD |
|---|---|---|---|---|---|---|
| **Defect** | 0.94 | 0.99 | 0.90 | 0.99 | 0.83 | 0.75 |
| | *p*-value < 0.001 | | *p*-value < 0.001 | | *p*-value < 0.03 | |
| **Extent** | 0.87 | 0.99 | 0.83 | 0.97 | 0.71 | 0.70 |
| | *p*-value < 0.001 | | *p*-value < 0.001 | | *p*-value < 0.01 | |
| **SSS / SRS** | 0.83 | 0.99 | 0.75 | 0.96 | 0.68 | 0.70 |
| | *p*-value < 0.001 | | *p*-value < 0.001 | | *p*-value < 0.01 | |
| **SS% / SR%** | 0.82 | 0.99 | 0.75 | 0.97 | 0.66 | 0.70 |
| | *p*-value < 0.001 | | *p*-value < 0.001 | | *p*-value < 0.01 | |
| **TPD%** | 0.88 | 0.99 | 0.82 | 0.98 | 0.76 | 0.76 |
| | *p*-value < 0.001 | | *p*-value < 0.001 | | *p*-value < 0.01 | |
| **Volume** | 0.99 | 1.00 | 0.99 | 1.00 | 0.95 | 0.97 |
| | *p*-value < 0.001 | | *p*-value < 0.03 | | *p*-value = 0.29 | |
| **Wall** | 0.98 | 1.00 | 0.97 | 1.00 | 0.93 | 0.96 |
| | *p*-value < 0.001 | | *p*-value < 0.001 | | *p*-value = 0.39 | |
| **Shape Eccentricity** | 0.94 | 0.99 | 0.93 | 0.99 | 0.84 | 0.82 |
| | *p*-value < 0.001 | | *p*-value < 0.001 | | *p*-value < 0.01 | |
| **Shape Index** | 0.72 | 0.95 | 0.60 | 0.88 | 0.54 | 0.72 |
| | *p*-value < 0.001 | | *p*-value < 0.004 | | *p*-value < 0.03 | |

HD: Half-Dose, PHD: Predicted Half-Dose, QD: Quarter-Dose, PQD: Predicted Quarter-Dose, OD: One-eighth-Dose, POD: Predicted One-eighth-Dose. *p*-value between the low-dose and the predicted normal-dose projections at each reduced dose level.

### 3.4. Clinical Evaluation

The summed scores assigned by the physician for the low-dose, normal-dose, and predicted normal-dose images, as well as diagnostic changes compared with the reference normal-dose, are presented in supplemental Table 1 for the entire patients in the external test dataset. Pearson correlation coefficients (Tabel 4) and bar plots (Figure 9) were employed to summarize the information in the supplemental Table 1. The Pearson correlation coefficients increased considerably for the scores assigned to the predicted normal-dose images compared to the corresponding low-dose images at the three dose levels. However, the prediction from one-eighth-dose exhibited a less significant correlation coefficient (86%). We carried out a paired sample t-test at the 5% significance level on the SS values distributions for the low-dose and the corresponding predicted normal-dose values. The t-test resulted in *p*-values < 0.001, which indicated that the mean values for the low-dose and predicted normal-dose data were statistically different.

A bar chart was used to display the diagnostic differences in the low-dose and predicted normal-dose images compared to the reference images (Figure 9). Figure 9 shows that the absolute value of the differences decreased from low-dose to predicted normal-dose at all three reduced dose levels. As a general conclusion based on the physician assessment, cases with the score difference of 0 and ±1 were considered clinically acceptable with no notable diagnostic changes which are indicated by hatched charts in Figure 9. In this light, the percentage of acceptable cases was 100% for the half-dose, 80% for the quarter-dose, and 11% at the quarter-dose level.

**Table 4.** Pearson correlation coefficients for SS values given by the nuclear medicine specialist. *p*-value between the low-dose and the predicted normal-dose projections at each reduced dose level.

| Parameter | HD | PHD | QD | PQD | OD | POD |
|---|---|---|---|---|---|---|
| **Pearson correlation coefficient** | 0.909 | 0.984 | 0.823 | 0.963 | 0.665 | 0.861 |
| | *p*-value < 0.001 | | *p*-value < 0.001 | | *p*-value < 0.001 | |

HD: Half-Dose, PHD: Predicted Half-Dose, QD: Quarter-Dose, PQD: Predicted Quarter-Dose, OD: One-eighth-Dose, POD: Predicted One-eighth-Dose.

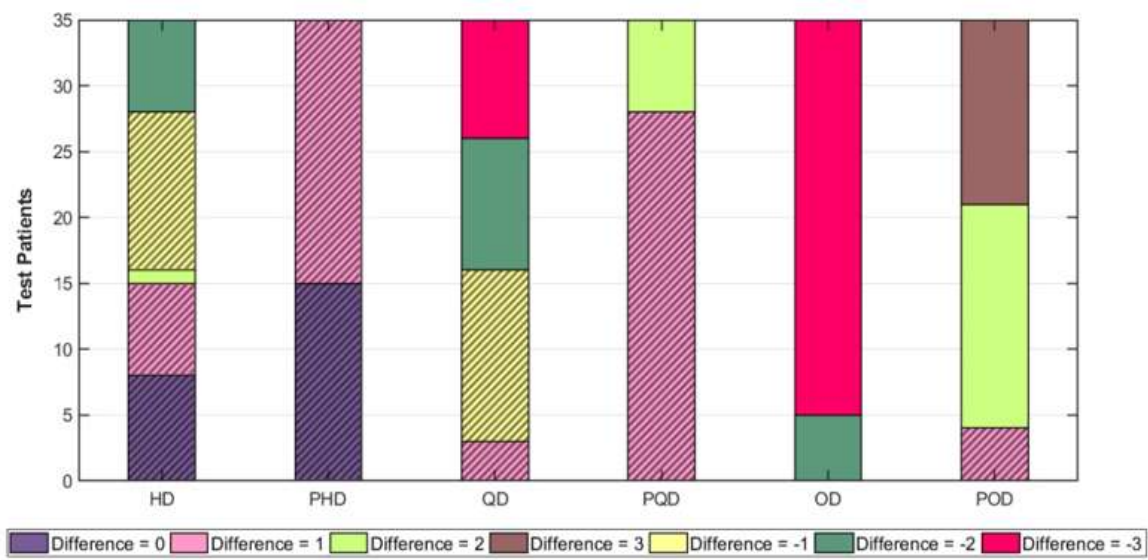

**Figure 9.** Result of image quality assessment (summed score difference) by the nuclear medicine specialist for the low-dose and predicted normal-dose images at the three reduced dose levels. Clinically acceptable cases are hatched. HD: Half-Dose, PHD: Predicted Half-Dose, QD: Quarter-Dose, PQD: Predicted Quarter-Dose, OD: One-eighth-Dose, POD: Predicted One-eighth-Dose.

## 4. Discussion

This study aimed to assess the possibility of dose reduction in SPECT-MPI without scarification of the diagnostic values of the resulting images. To this end, a generative adversarial network was employed to predict high-quality normal-dose SPECT images from the corresponding low-dose data in the projection space. The proposed network was applied to suppress noise in the non-gated projections at the different reduced dose levels. Clinical assessment showed that the proposed network has a promising performance for the half- and quarter-dose levels. The entire estimated data at the half-dose level were considered clinically acceptable; however, at the quarter-dose level, this number was reduced to 80%. Evaluation at the quarter-dose level revealed that almost all of the poor quality predicted cases are associated with the patients who were diagnosed with moderate/severe-risk conditions. Dose reduction in gated SPECT-MPI was also considered, however since the signal-to-noise ratio (SNR) in gated images is remarkably poor compared to the nongated imaging only half-dose level was studied. The fact that gated imaging was conducted in 8-time intervals, the projection data already bore high noise levels and dose reduction by half led to extremely poor SNR. Figure 10 shows a representative predicted gated projection compared to the half-dose and the reference normal-dose counterparts, wherein the excessive amount of dose in the reduced dose, as well as the reference gated projections, led to over-smoothed prediction. Dose reduction in gated SPECT-MPI faces the challenge of poor SNR and noise-induced artifacts, however deep learning solutions (such as the proposed GAN model in this study) could be employed to enhance the quality of normal/conventional gated SPECT-MPI (without dose reduction) as they almost have the same signal to noise properties of the one-eight low-dose nongated images.

Regarding similar works in SPECT-MPI, Ramon et al. [25] proposed a couple of 3D convolutional auto-encoders with/without skip connections to denoise the reconstructed non-gated images (corrected for attenuation and scatter) at 1/2, 1/4, 1/8, and 1/16 of standard-dose levels. Reconstruction strategies used in this study were optimized previously for low-dose acquisitions in [14]. They reported a Pearson correlation coefficient of $0.992 \pm 0.001$ for predicted images compared to $0.982 \pm 0.001$ for the low-dose input at the 1/4-dose level with respect to normal-dose OSEM reconstructed images. A similar observation was made in our study wherein the Pearson correlation coefficient improved from $\rho = 0.987 \pm 0.007$ to $\rho = 0.994 \pm 0.003$ at the quarter-dose. It should be noted Ramon et al. performed the data acquisition using a one-day rest/stress SPECT MPI protocol with administrated activity doses ranging from 30 mCi to 36 mCi, while in our study, the patients underwent a two-day rest/stress SPECT MPI protocol with administrated activity doses between 20 and 25 mCi. Therefore, the absolute amount of injected dose at the quarter-dose level in our study is less than the administrated dose in the above-mentioned study, nevertheless, a significant/comparable improvement was observed at the quarter-dose level.

Shiri et al. [27] employed a residual deep learning model to reduce the acquisition time per projection by 50%, wherein the full-time projections were predicted from the corresponding half-time projections. The reconstruction of the predicted projections led to SSIM = 0.98 ± 0.01, PSNR = 36.0 ± 1.4, RMSE = 3.1 ± 1.1, and (Pearson correlation) $\rho$ = 0.987 in the image domain. Shiri et al. only studied half-time acquisition in their work (they did not investigate 1/4- and 1/8-dose levels neither gated imaging), where relatively better performance was observed at 1/2-dose level in our study with SSIM of 0.99 ± 0.01, PSNR of 42.49 ± 2.37, RMSE of 1.99 ± 0.63, and Pearson correlation coefficient of 0.997 ± 0.001. It should be noted that both studies used the same SPECT scanner as well as acquisition (injected dose) protocols. In this study, in addition to the investigation of the 1/4-dose and 1/8-dose levels, a clinical assessment was performed by a nuclear medicine specialist to provide a useful insight into the clinical values of the resulting images.

Though there is a fundamental difference between fast image acquisition and low dose imaging, the reduced number of detected photons would lead to lower SNRs in both scenarios. However fast image acquisition would be less affected by the involuntary patient motion such as respiratory motion. In this light, the noise suppression model investigated in this study could also be employed for fast image acquisition protocols.

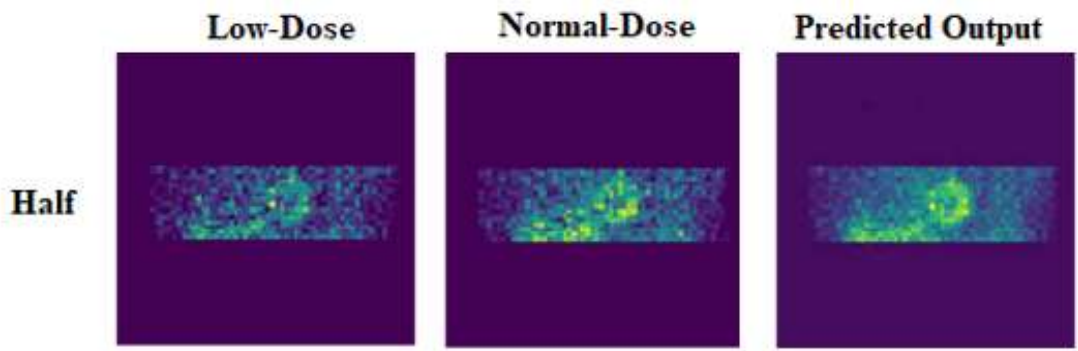

**Figure 10.** Illustration of the predicted gated projections for a randomly selected patient from the test dataset at the half-dose level compared to the reference normal-dose and low-dose projections.

## 5. Conclusion

This study set out to investigate the feasibility of the dose reduction in SPECT-MPI without sacrificing the quantitative accuracy and clinical value of the resulting images. A deep learning solution was assessed at different reduced dose levels to predict standard projection data from low-dose counterparts. The quantitative analysis, as well as the clinical assessment, demonstrated that the deep learning model could effectively recover the underlying information in 1/2-dose and 1/4-dose SPECT images. However, due to the extremely high noise levels in 1/8-dose and gated SPECT-MPI, the deep learning model failed to fully recover the underlying signals/image quality, which warrants further investigation.

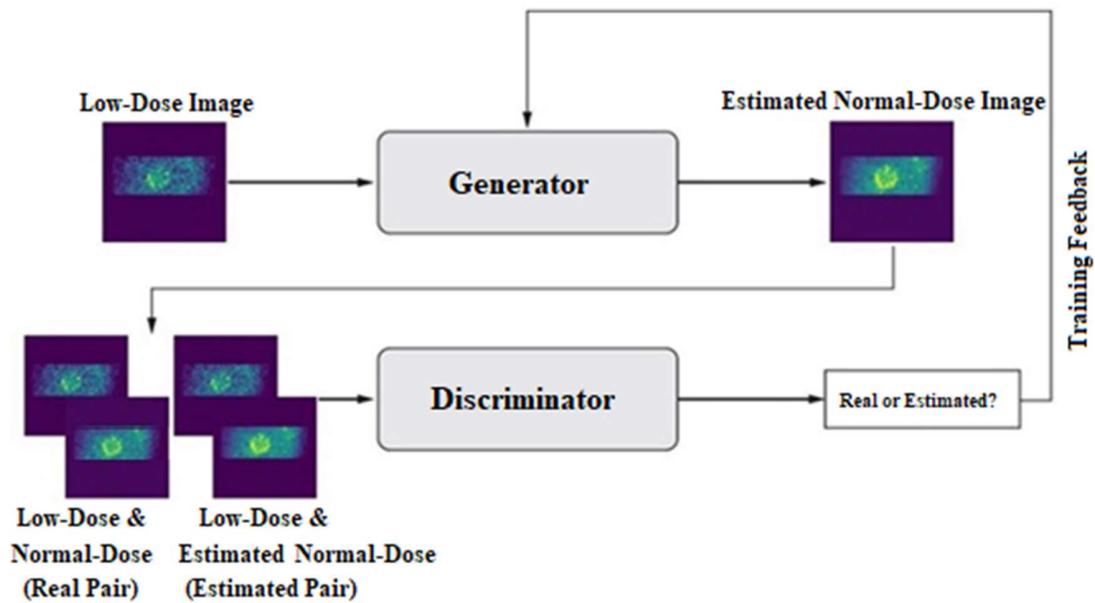

**Supplemental Figure 1.** Schematic illustration of the implemented GAN model.

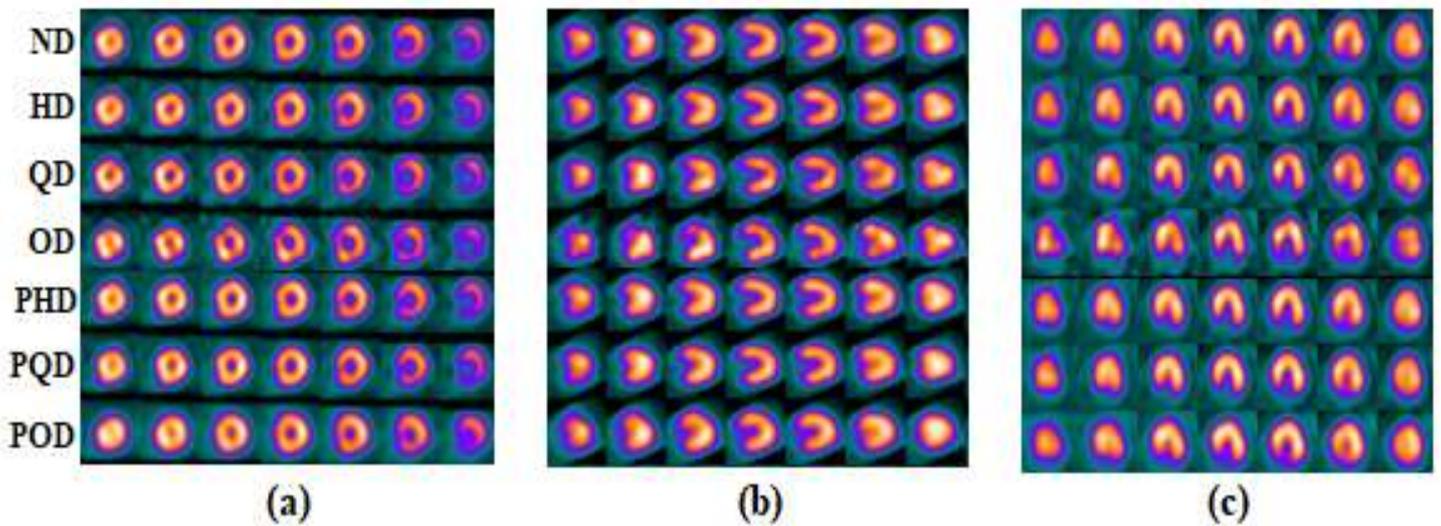

**Supplemental Figure 2**. Reconstructed non-gated images for a patient diagnosed with normal perfusion. (a): short-axis view, (b): long vertical-axis view, (c): horizontal long-axis view. In (a), (b), and (c), the rows from top to bottom correspond to the normal-dose (ND), half-dose (HD), quarter-dose (QD), one-eighth-dose (OD), predicted half-dose (PHD), predicted quarter dose (PQD), and predicted one-eighth-dose (POD), respectively.

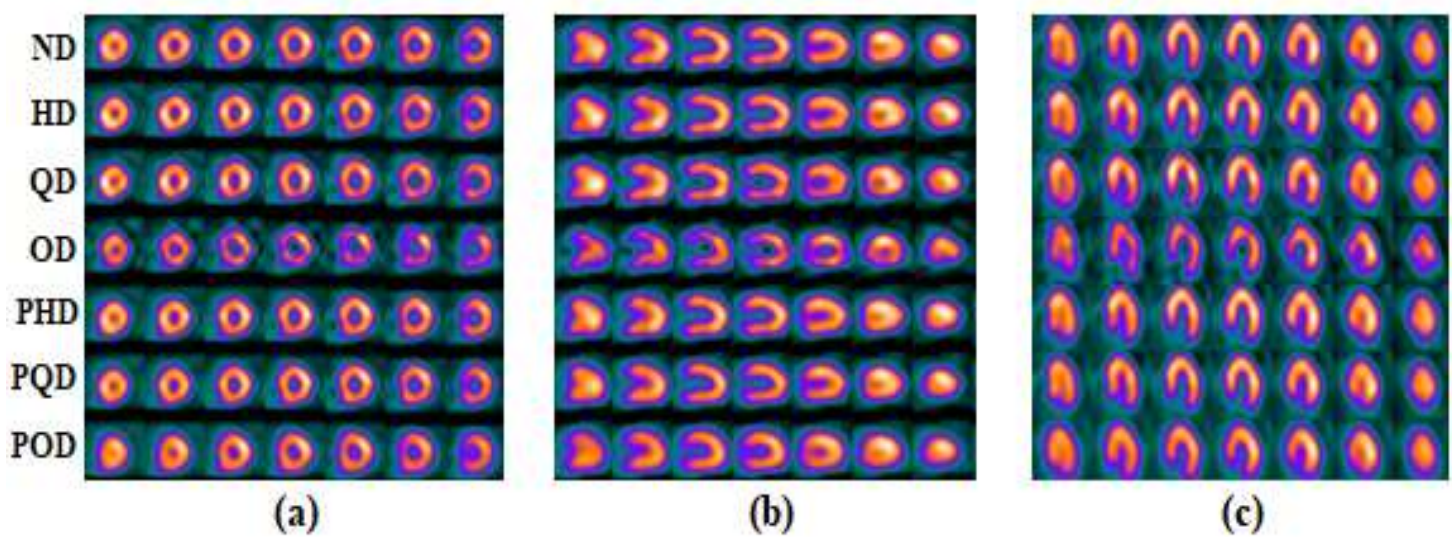

**Supplemental Figure 3**. Reconstructed non-gated images for a patient diagnosed with low-risk. (a): short-axis view, (b): long vertical-axis view, (c): horizontal long-axis view. In (a), (b), and (c), the rows from top to bottom correspond to the normal-dose (ND), half-dose (HD), quarter-dose (QD), one-eighth-dose (OD), predicted half-dose (PHD), predicted quarter dose (PQD), and predicted one-eighth-dose (POD), respectively.

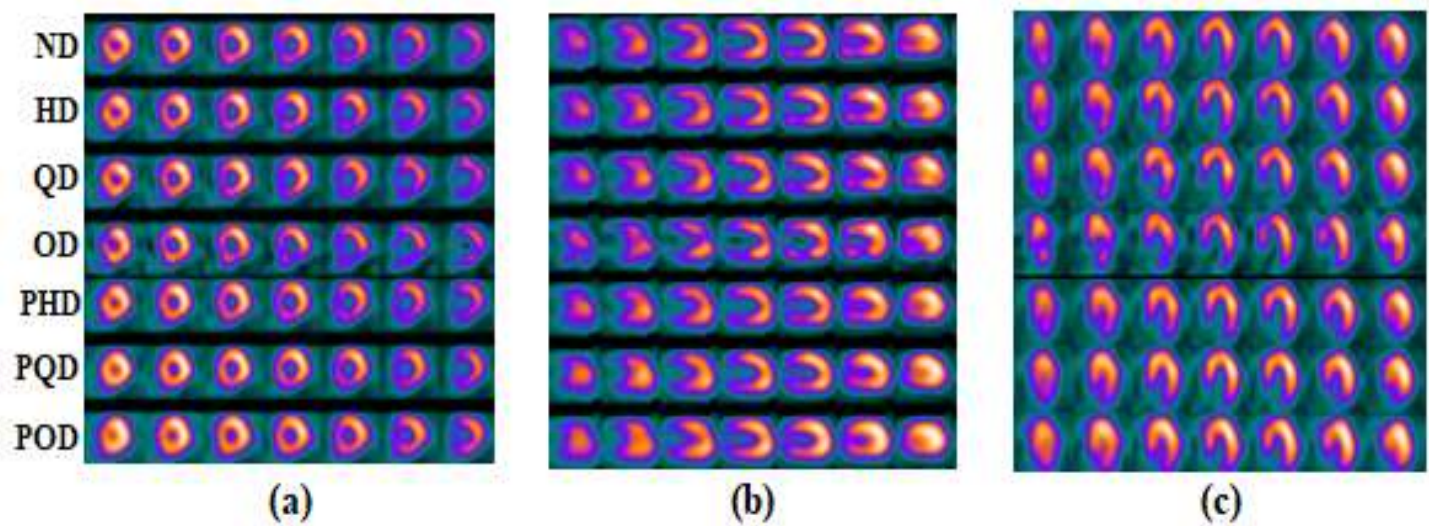

**Supplemental Figure 4**. Reconstructed non-gated images for a patient diagnosed with intermediate-risk. (a): short-axis view, (b): long vertical-axis view, (c): horizontal long-axis view. In (a), (b), and (c), the rows from top to bottom correspond to the normal-dose (ND), half-dose (HD), quarter-dose (QD), one-eighth-dose (OD), predicted half-dose (PHD), predicted quarter dose (PQD), and predicted one-eighth-dose (POD), respectively.

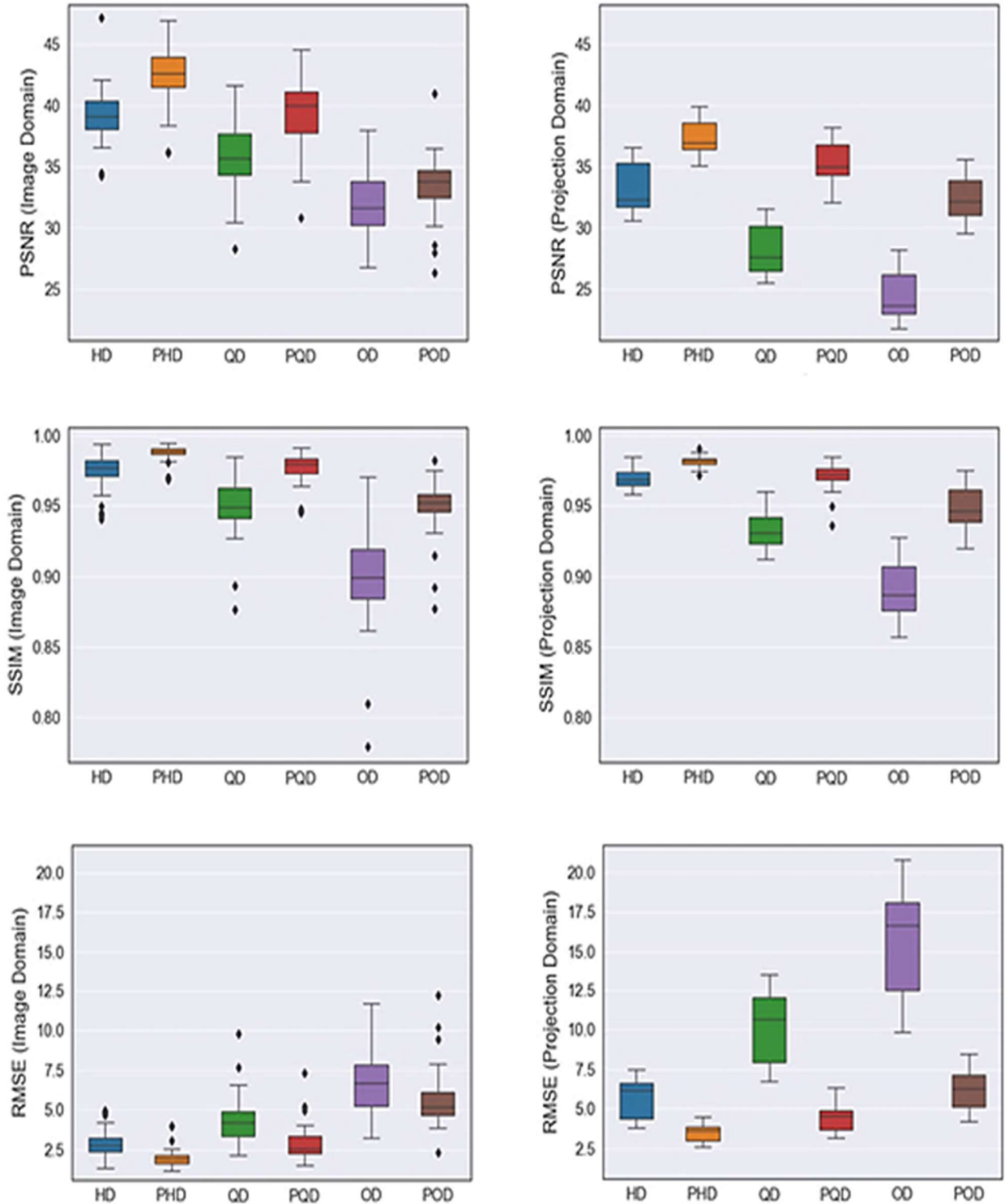

**Supplemental Figure 5.** Box plots comparing the quantitative parameters in the image and projection spaces. The rows from top to bottom correspond to PSNR, SSIM, and RMSE, respectively. The left and right columns show the results of the image domain and projection domains, respectively. HD: Half-Dose, PHD: Predicted Half-Dose, QD: Quarter-Dose, PQD: Predicted Quarter-Dose, OD: One-eighth-Dose, POD: Predicted One-eighth-Dose.

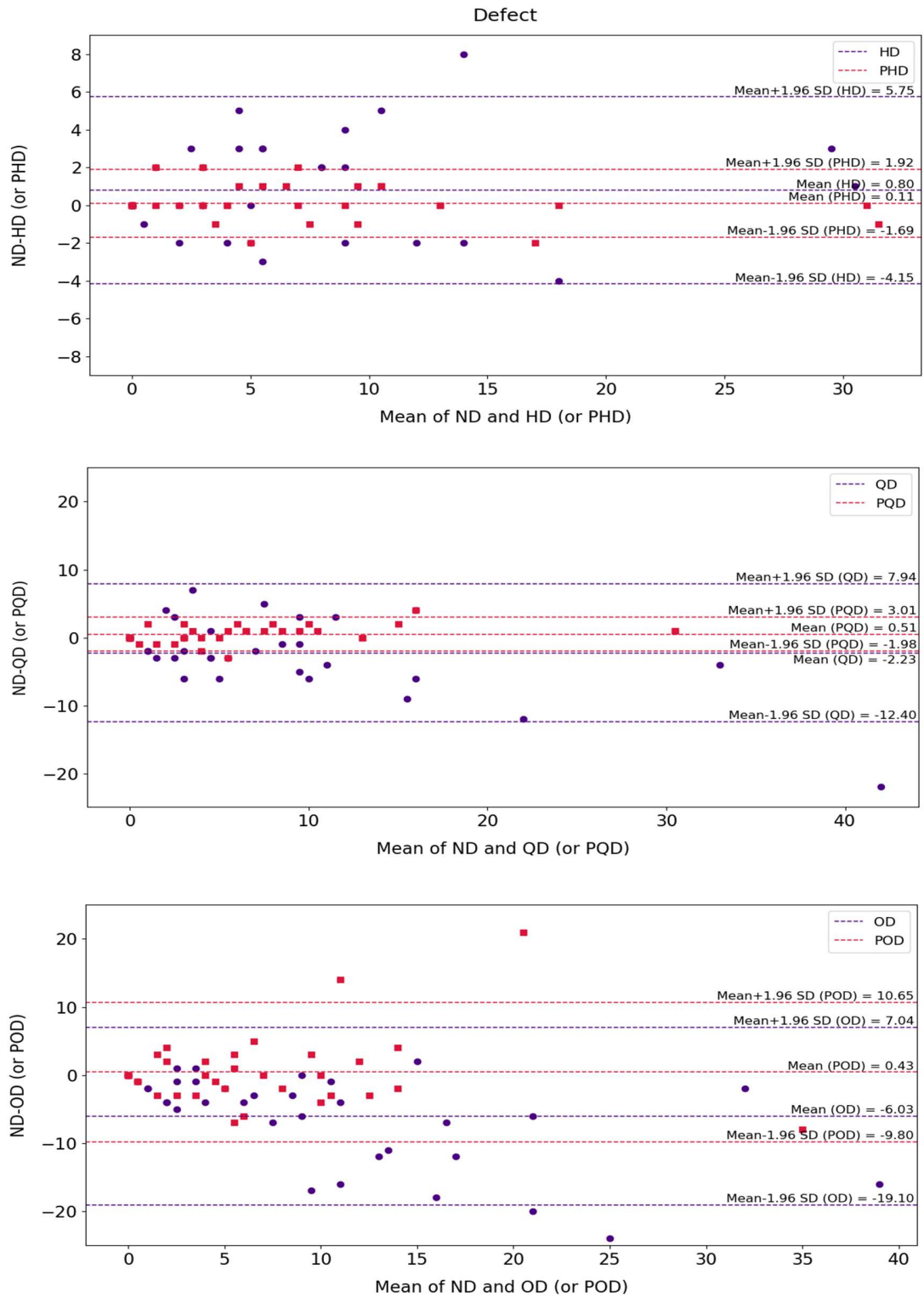

**Supplemental Figure 6.** Bland-Altman plots of Defect index for the low-dose and predicted normal-dose images at (a) half-dose level, (b) quarter-dose levels, and (c) one-eighth-dose level compared with the reference normal-dose images. The blue and red dashed lines designate the mean and 95% confidence interval of the Defect differences in the low-dose and predicted normal-dose images, respectively. HD: Half-Dose, PHD: Predicted Half-Dose, QD: Quarter-Dose, PQD: Predicted Quarter-Dose, OD: One-eighth-Dose, POD: Predicted One-eighth-Dose.

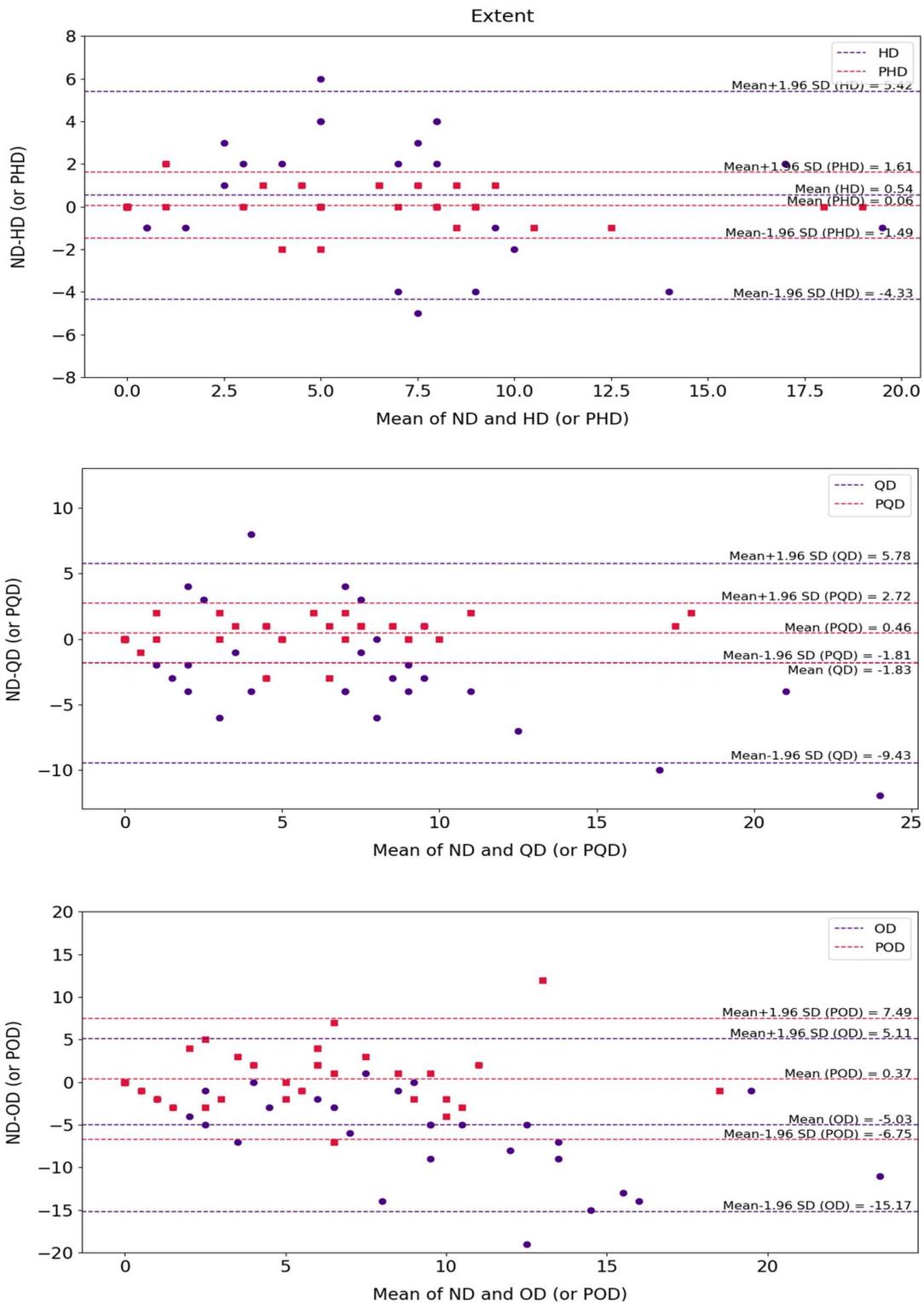

**Supplemental Figure 7.** Bland-Altman plots of Extent index for the low-dose and predicted normal-dose images at (a) half-dose level, (b) quarter-dose levels, and (c) one-eighth-dose level compared with the reference normal-dose images. The blue and red dashed lines designate the mean and 95% confidence interval of the Extent differences in the low-dose and predicted normal-dose images, respectively. HD: Half-Dose, PHD: Predicted Half-Dose, QD: Quarter-Dose, PQD: Predicted Quarter-Dose, OD: One-eighth-Dose, POD: Predicted One-eighth-Dose.

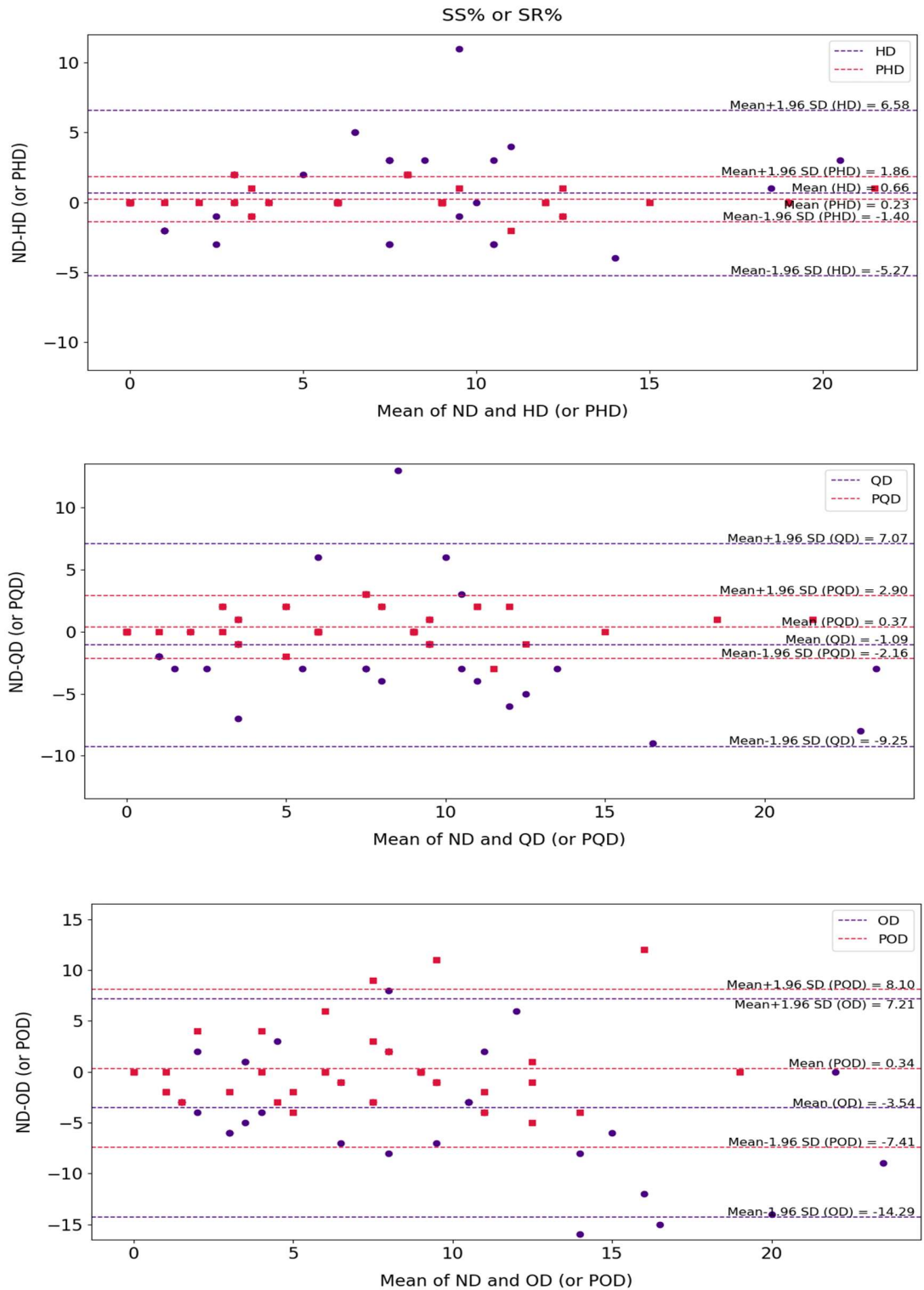

**Supplemental Figure 8.** Bland-Altman plots of SS% / SR% index for the low-dose and predicted normal-dose images at (a) half-dose level, (b) quarter-dose level, and (c) one-eighth-dose levels compared with the reference normal-dose images. The blue and red dashed lines designate the mean and 95% confidence interval of the SS% / SR% differences in the low-dose and predicted normal-dose images, respectively. HD: Half-Dose, PHD: Predicted Half-Dose, QD: Quarter-Dose, PQD: Predicted Quarter-Dose, OD: One-eighth-Dose, POD: Predicted One-eighth-Dose.

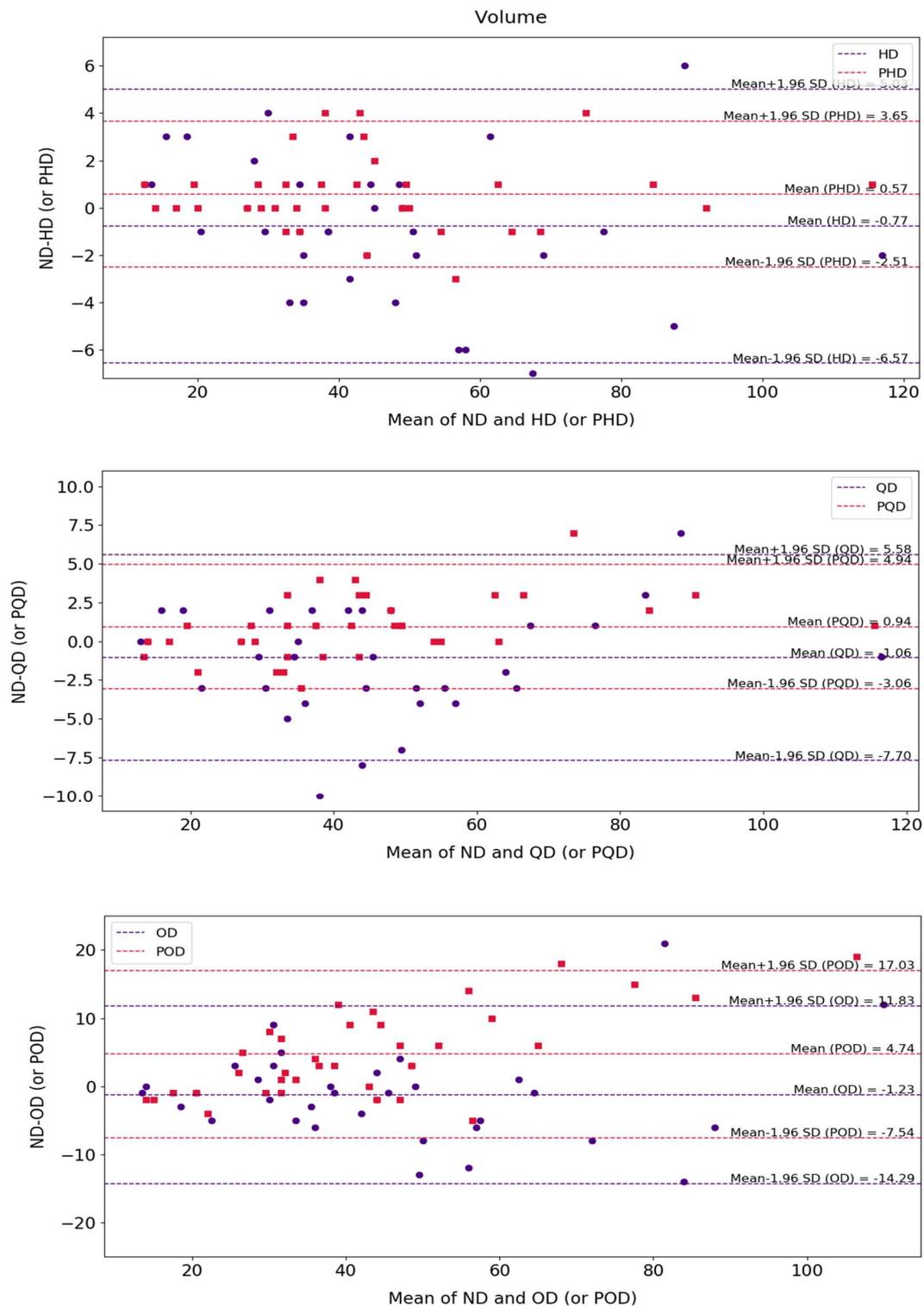

**Supplemental Figure 9.** Bland-Altman plots of Volume index for the low-dose and predicted normal-dose images at (a) half-dose, (b) quarter-dose, and (c) one-eighth-dose levels compared with the reference normal-dose images. The blue and red dashed lines designate the mean and 95% confidence interval of the Volume differences in the low-dose and predicted normal-dose images, respectively. HD: Half-Dose, PHD: Predicted Half-Dose, QD: Quarter-Dose, PQD: Predicted Quarter-Dose, OD: One-eighth-Dose, POD: Predicted One-eighth-Dose.

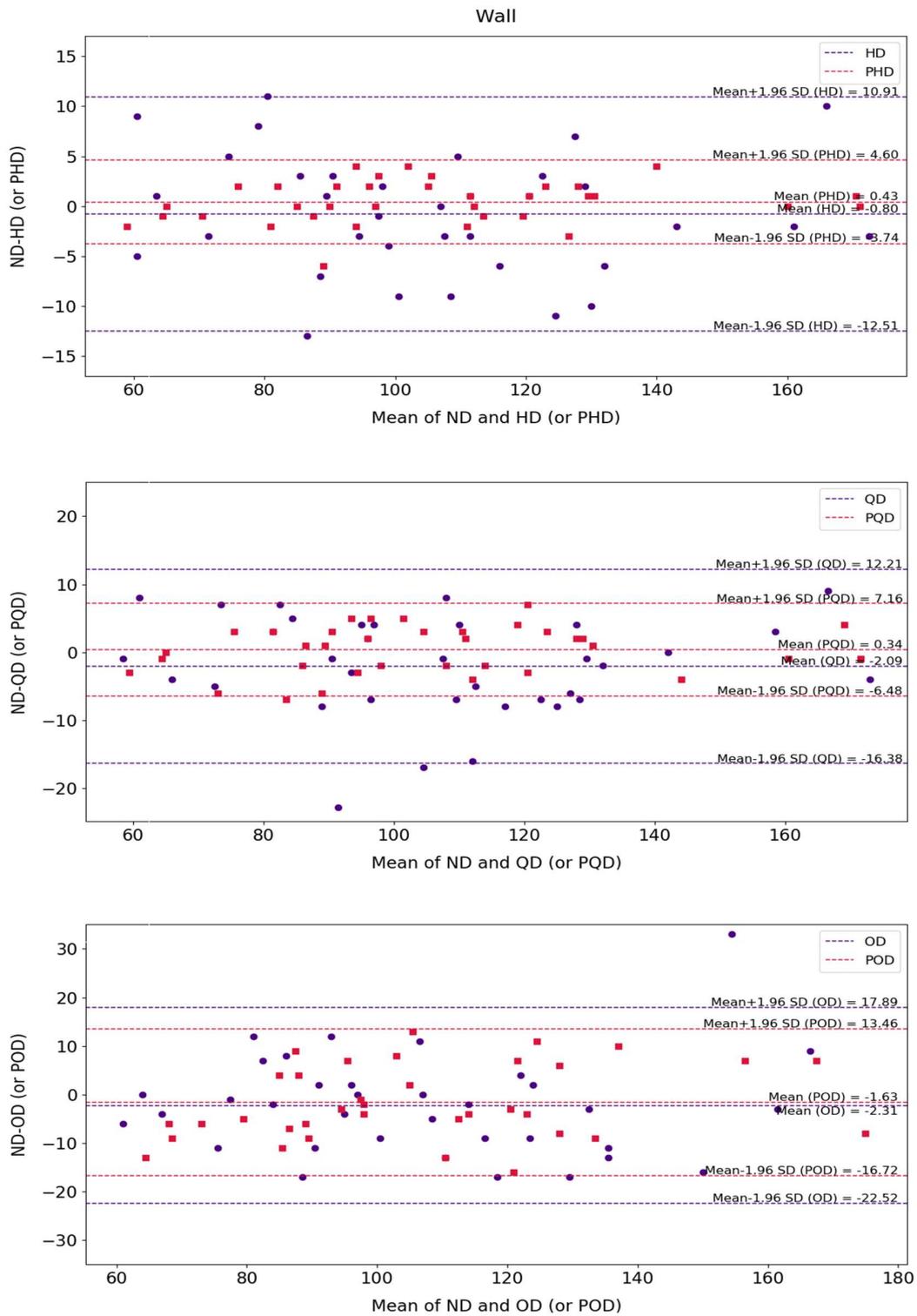

**Supplemental Figure 10.** Bland-Altman plots of Wall index for the low-dose and predicted normal-dose images at (a) half-dose, (b) quarter-dose, and (c) one-eighth-dose levels compared with the reference normal-dose images. The blue and red dashed lines designate the mean and 95% confidence interval of the Wall differences in the low-dose and predicted normal-dose images, respectively. HD: Half-Dose, PHD: Predicted Half-Dose, QD: Quarter-Dose, PQD: Predicted Quarter-Dose, OD: One-eighth-Dose, POD: Predicted One-eighth-Dose.

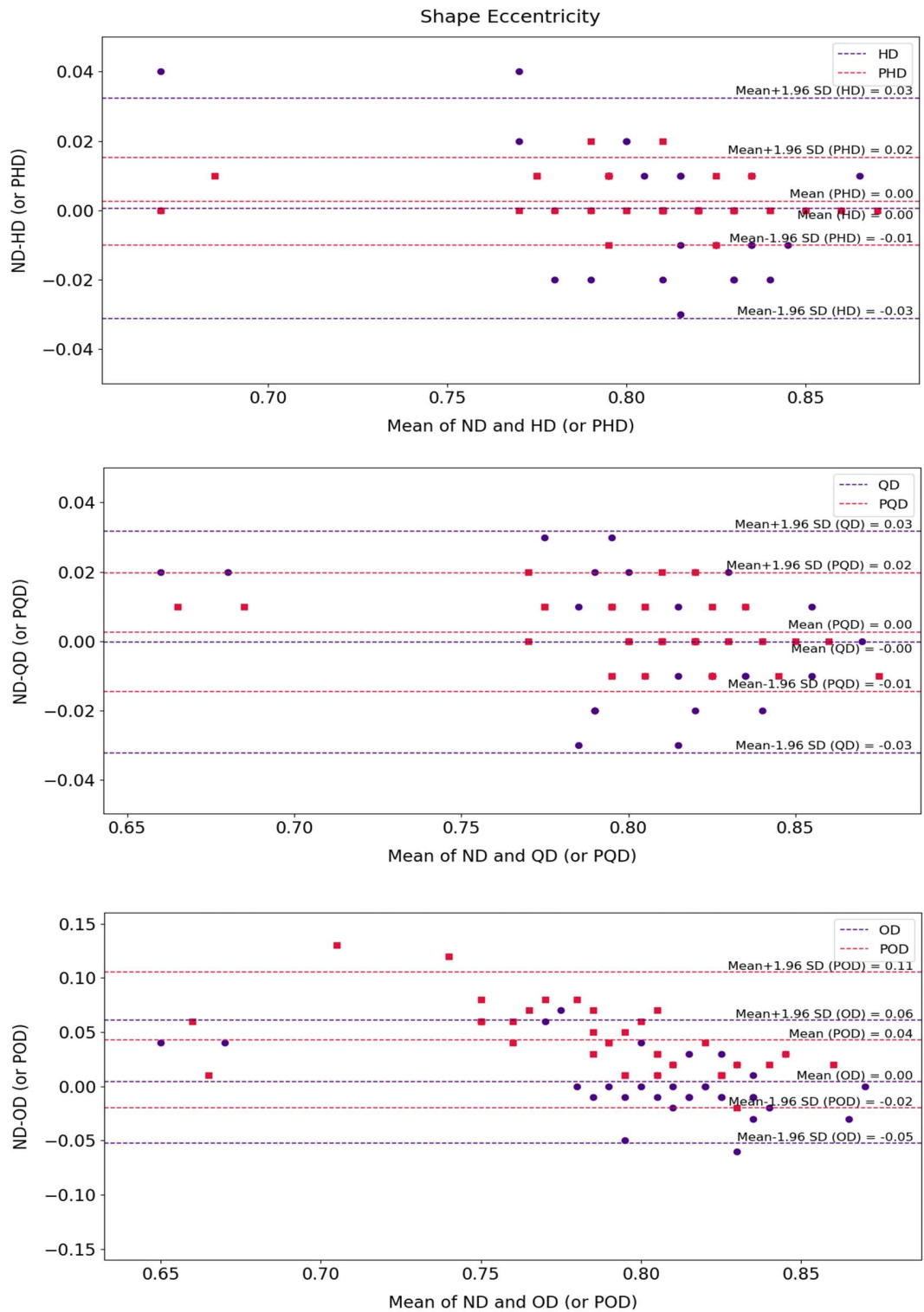

**Supplemental Figure 11.** Bland-Altman plots of Shape Eccentricity index for the low-dose and predicted normal-dose images at (a) half-dose, (b) quarter-dose, and (c) one-eighth-dose levels compared with the reference normal-dose images. The blue and red dashed lines designate the mean and 95% confidence interval of the Shape Eccentricity differences in the low-dose and predicted normal-dose images, respectively. HD: Half-Dose, PHD: Predicted Half-Dose, QD: Quarter-Dose, PQD: Predicted Quarter-Dose, OD: One-eighth-Dose, POD: Predicted One-eighth-Dose.

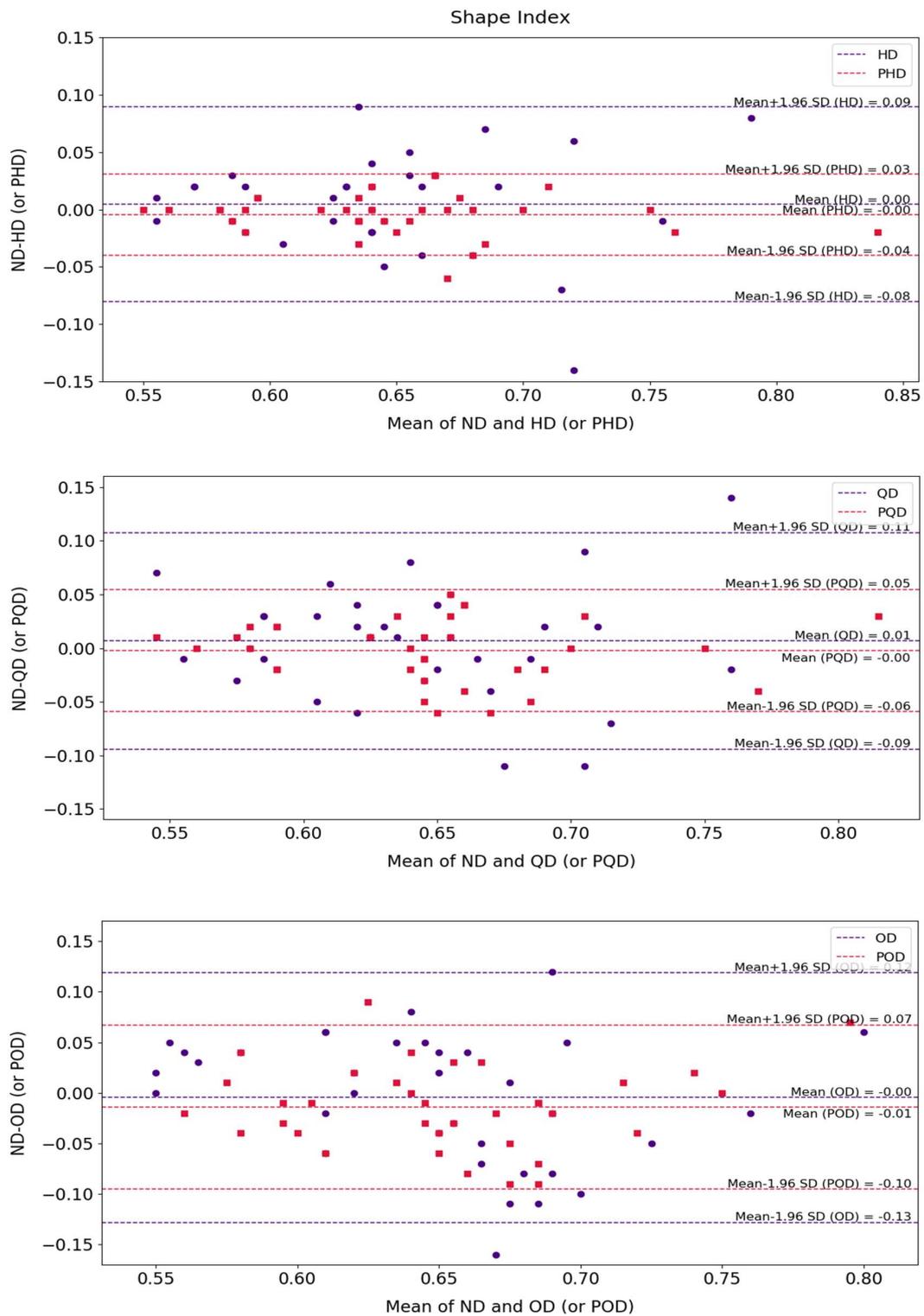

**Supplemental Figure 12.** Bland-Altman plots of Shape Index for the low-dose and predicted normal-dose images at (a) half-dose, (b) quarter-dose, and (c) one-eighth-dose levels compared with the reference normal-dose images. The blue and red dashed lines designate the mean and 95% confidence interval of the Shape Index differences in the low-dose and predicted normal-dose images, respectively. HD: Half-Dose, PHD: Predicted Half-Dose, QD: Quarter-Dose, PQD: Predicted Quarter-Dose, OD: One-eighth-Dose, POD: Predicted One-eighth-Dose.

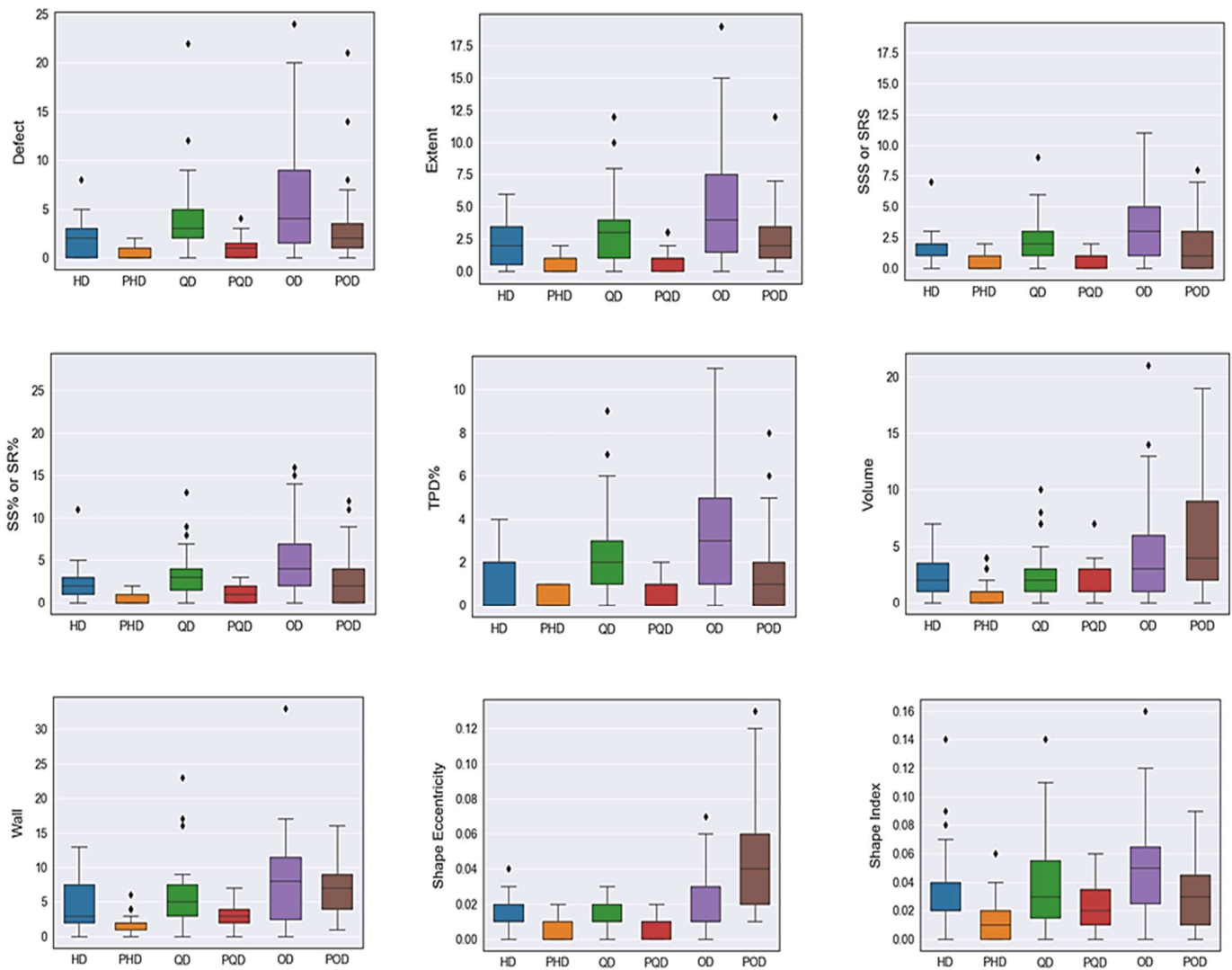

**Supplemental Figure 13.** The box plots of the Defect, Extent, SS% or SR%, SSS or SRS, TPD%, Volume, Wall, Shape Eccentricity, and Shape Index. Absolute differences between the low-dose/predicted normal-dose images and the reference normal-dose data at half-dose, quarter-dose, and one-eighth-dose levels. HD: Half-Dose, PHD: Predicted Half-Dose, QD: Quarter-Dose, PQD: Predicted Quarter-Dose, OD: One-eighth-Dose, POD: Predicted One-eighth-Dose.

**Supplemental Table 1.** Summed score (SS) values assigned by the nuclear medicine specialist for 35 patients in the test dataset for the non-gated normal, half, quarter, and one-eighth dose levels. Difference (Dif) values show the diagnostic changes in the low-dose/predicted normal-dose images compared to the reference normal-dose data at the different dose levels.

| Patients | ND | HD | | PHD | | QD | | PQD | | OD | | POD | |
|---|---|---|---|---|---|---|---|---|---|---|---|---|---|
| | SS | SS | Dif | SS | Dif | SS | Dif | SS | Dif | SS | Dif | SS | Dif |
| 1 | 6 | 7 | -1 | 6 | 0 | 10 | -2 | 3 | +1 | 15 | -3 | 0 | +2 |
| 2 | 18 | 22 | -2 | 15 | +1 | 32 | -3 | 12 | +2 | 48 | -3 | 9 | +3 |
| 3 | 14 | 16 | -1 | 12 | +1 | 20 | -3 | 9 | +2 | 22 | -3 | 7 | +2 |
| 4 | 3 | 8 | -2 | 3 | 0 | 16 | -3 | 2 | +1 | 30 | -3 | 0 | +2 |
| 5 | 1 | 2 | -1 | 0 | +1 | 2 | -1 | 0 | +1 | 6 | -2 | 0 | +1 |
| 6 | 3 | 4 | -1 | 2 | +1 | 8 | -2 | 2 | +1 | 6 | -2 | 0 | +3 |
| 7 | 7 | 5 | +1 | 5 | +1 | 11 | -2 | 5 | +1 | 24 | -3 | 1 | +3 |
| 8 | 5 | 6 | -1 | 5 | 0 | 8 | -1 | 4 | +1 | 14 | -3 | 2 | +3 |
| 9 | 23 | 22 | +1 | 20 | +1 | 28 | -2 | 17 | +2 | 41 | -3 | 11 | +3 |
| 10 | 5 | 3 | +1 | 4 | +1 | 11 | -3 | 4 | +1 | 21 | -3 | 2 | +2 |
| 11 | 6 | 6 | 0 | 6 | 0 | 8 | -1 | 5 | +1 | 22 | -3 | 2 | +2 |
| 12 | 12 | 8 | +2 | 11 | +1 | 10 | +1 | 7 | +2 | 25 | -3 | 3 | +3 |
| 13 | 14 | 20 | -2 | 12 | +1 | 22 | -3 | 10 | +2 | 26 | -3 | 2 | +3 |
| 14 | 3 | 3 | 0 | 3 | 0 | 8 | -2 | 2 | +1 | 14 | -3 | 0 | +2 |
| 15 | 4 | 4 | 0 | 4 | 0 | 6 | -1 | 3 | +1 | 13 | -3 | 1 | +3 |
| 16 | 4 | 3 | +1 | 4 | 0 | 8 | -1 | 1 | +2 | 22 | -3 | 0 | +3 |
| 17 | 4 | 4 | 0 | 4 | +1 | 6 | -1 | 3 | +1 | 12 | -2 | 1 | +3 |
| 18 | 8 | 6 | +1 | 6 | +1 | 6 | +1 | 3 | +2 | 32 | -3 | 6 | +1 |
| 19 | 4 | 4 | 0 | 3 | +1 | 6 | -1 | 2 | +1 | 18 | -3 | 1 | +2 |
| 20 | 3 | 9 | -2 | 3 | 0 | 11 | -3 | 2 | +1 | 26 | -3 | 0 | +2 |
| 21 | 10 | 12 | -1 | 8 | +1 | 19 | -3 | 7 | +1 | 38 | -3 | 4 | +2 |
| 22 | 4 | 6 | -1 | 3 | +1 | 12 | -3 | 3 | +1 | 24 | -3 | 1 | +2 |
| 23 | 8 | 12 | -2 | 7 | +1 | 18 | -3 | 6 | +1 | 32 | -3 | 3 | +3 |
| 24 | 4 | 5 | -1 | 4 | 0 | 6 | -1 | 3 | +1 | 18 | -3 | 1 | +2 |
| 25 | 5 | 4 | +1 | 3 | +1 | 7 | -1 | 4 | +1 | 22 | -3 | 1 | +2 |
| 26 | 4 | 4 | 0 | 3 | +1 | 6 | -1 | 3 | +1 | 14 | -3 | 2 | +2 |
| 27 | 12 | 14 | -1 | 10 | +1 | 16 | -2 | 9 | +1 | 33 | -3 | 4 | +3 |
| 28 | 7 | 8 | -1 | 5 | +1 | 14 | -2 | 5 | +1 | 35 | -3 | 3 | +2 |
| 29 | 3 | 3 | 0 | 3 | 0 | 2 | +1 | 2 | +1 | 6 | -2 | 1 | +2 |
| 30 | 4 | 3 | +1 | 4 | 0 | 6 | -1 | 3 | +1 | 16 | -3 | 1 | +2 |
| 31 | 2 | 4 | -1 | 0 | +1 | 8 | -2 | 0 | +1 | 14 | -3 | 1 | +1 |

| | ND | HD | PHD | QD | PQD | OD | POD | | | | | | |
|---|---|---|---|---|---|---|---|---|---|---|---|---|---|
| 32 | 7 | 11 | -2 | 7 | 0 | 12 | -2 | 5 | +1 | 25 | -3 | 3 | +2 |
| 33 | 2 | 6 | -2 | 2 | 0 | 14 | -2 | 1 | +1 | 28 | -3 | 1 | +1 |
| 34 | 14 | 15 | -1 | 14 | 0 | 16 | -1 | 12 | +1 | 21 | -2 | 6 | +3 |
| 35 | 11 | 11 | 0 | 11 | 0 | 14 | -1 | 8 | +1 | 22 | -3 | 6 | +3 |

ND: Normal-Dose, HD: Half-Dose, PHD: Predicted Half-Dose, QD: Quarter-Dose, PQD: Predicted Quarter-Dose, OD: One-eighth-Dose, POD: Predicted One-eighth-Dose.

# References

[1] Gullberg GT, Reutter BW, Sitek A, Maltz JS, Budinger TF. Dynamic single photon emission computed tomography--basic principles and cardiac applications. Physics in medicine and biology. 2010;55:R111-91.

[2] Sajedi S, Zeraatkar N, Moji V, Farahani MH, Sarkar S, Arabi H, et al. Design and development of a high resolution animal SPECT scanner dedicated for rat and mouse imaging. Nuclear Instruments and Methods in Physics Research Section A: Accelerators, Spectrometers, Detectors and Associated Equipment. 2014;741:169-76.

[3] Cassar A, Holmes DR, Jr., Rihal CS, Gersh BJ. Chronic coronary artery disease: diagnosis and management. Mayo Clinic proceedings. 2009;84:1130-46.

[4] Mostafapour S, Gholamiankhah F, Maroofpour S, Momennezhad M, Asadinezhad M, Zakavi SR, et al. Deep learning-based attenuation correction in the image domain for myocardial perfusion SPECT imaging. arXiv preprint arXiv:210204915. 2021.

[5] Wells RG. Dose reduction is good but it is image quality that matters. Journal of nuclear cardiology : official publication of the American Society of Nuclear Cardiology. 2020;27:238-40.

[6] Sanaat A, Arabi H, Mainta I, Garibotto V, Zaidi H. Projection Space Implementation of Deep Learning-Guided Low-Dose Brain PET Imaging Improves Performance over Implementation in Image Space. Journal of nuclear medicine. 2020;61:1388-96.

[7] Food U, Administration D. Initiative to reduce unnecessary radiation exposure from medical imaging. Center for Devices and Radiological Health, ed. 2010.

[8] Jerome SD, Tilkemeier PL, Farrell MB, Shaw LJ. Nationwide Laboratory Adherence to Myocardial Perfusion Imaging Radiation Dose Reduction Practices: A Report From the Intersocietal Accreditation Commission Data Repository. JACC Cardiovascular imaging. 2015;8:1170-6.

[9] Mostafapour S, Arabi H, Gholamian Khah F, Razavi-Ratki SK, Parach AA. Tc-99m (methylene diphosphonate) SPECT quantitative imaging: Impact of attenuation map generation from SPECT-non-attenuation corrected and MR images on diagnosis of metastatic bone INTERNATIONAL JOURNAL OF RADIATION RESEARCH. 2021;In press.

[10] Khoshyari-morad Z, Jahangir R, Miri-Hakimabad H, Mohammadi N, Arabi H. Monte Carlo-based estimation of patient absorbed dose in 99mTc-DMSA, -MAG3, and -DTPA SPECT imaging using the University of Florida (UF) phantoms. arXiv:210300619. 2021.

[11] Arabi H, Zaidi H. Applications of artificial intelligence and deep learning in molecular imaging and radiotherapy. Euro J Hybrid Imaging. 2020;4(1), 1:23.

[12] Arabi H, Zaidi H. Spatially guided nonlocal mean approach for denoising of PET images. Medical physics. 2020;47:1656-69.

[13] Case JA. 3D iterative reconstruction can do so much more than reduce dose. Journal of nuclear cardiology : official publication of the American Society of Nuclear Cardiology. 2019.

[14] Juan Ramon A, Yang Y, Pretorius PH, Slomka PJ, Johnson KL, King MA, et al. Investigation of dose reduction in cardiac perfusion SPECT via optimization and choice of the image reconstruction strategy. Journal of nuclear cardiology : official publication of the American Society of Nuclear Cardiology. 2018;25:2117-28.

[15] Juan Ramon A, Yang Y, Wernick MN, Pretorius PH, Johnson KL, Slomka PJ, et al. Evaluation of the effect of reducing administered activity on assessment of function in cardiac gated SPECT. Journal of nuclear cardiology : official publication of the American Society of Nuclear Cardiology. 2020;27:562-72.

[16] Zeraatkar N, Sajedi S, Farahani MH, Arabi H, Sarkar S, Ghafarian P, et al. Resolution-recovery-embedded image reconstruction for a high-resolution animal SPECT system. Physica medica : PM : an international journal devoted to the applications of physics to medicine and biology : official journal of the Italian Association of Biomedical Physics (AIFB). 2014;30:774-81.

[17] Arabi H, Zaidi H. Improvement of image quality in PET using post-reconstruction hybrid spatial-frequency domain filtering. Physics in medicine and biology. 2018;63:215010.

[18] Arabi H, Zaidi H. Non-local mean denoising using multiple PET reconstructions. Annals of nuclear medicine. 2020.

[19] Liu H, Wu J, Lu W, Onofrey JA, Liu YH, Liu C. Noise reduction with cross-tracer and cross-protocol deep transfer learning for low-dose PET. Physics in medicine and biology. 2020;65:185006.

[20] Gholizadeh-Ansari M, Alirezaie J, Babyn P. Deep Learning for Low-Dose CT Denoising Using Perceptual Loss and Edge Detection Layer. Journal of digital imaging. 2020;33:504-15.

[21] Ding Q, Chen G, Zhang X, Huang Q, Ji H, Gao H. Low-dose CT with deep learning regularization via proximal forward-backward splitting. Physics in medicine and biology. 2020;65:125009.

[22] Zhou L, Schaefferkoetter JD, Tham IWK, Huang G, Yan J. Supervised learning with cyclegan for low-dose FDG PET image denoising. Medical image analysis. 2020;65:101770.

[23] Ghane B, Karimian A, Mostafapour S, Gholamiankhak F, Shojaerazavi S, Arabi H. Quantitative analysis of image quality in low-dose CT imaging for Covid-19 patients. arXiv preprint arXiv:210208128. 2021.

[24] Sanaei B, Faghihi R, Arabi H. Quantitative investigation of low-dose PET imaging and post-reconstruction smoothing. arXiv:210310541. 2021.

[25] Ramon AJ, Yang Y, Pretorius PH, Johnson KL, King MA, Wernick MN. Improving Diagnostic Accuracy in Low-Dose SPECT Myocardial Perfusion Imaging With Convolutional Denoising Networks. IEEE transactions on medical imaging. 2020;39:2893-903.

[26] Song C, Yang Y, Wernick MN, Pretorius PH, King MA. Low-Dose Cardiac-Gated Spect Studies Using a Residual Convolutional Neural Network.  2019 IEEE 16th International Symposium on Biomedical Imaging (ISBI 2019)2019. p. 653-6.

[27] Shiri I, AmirMozafari Sabet K, Arabi H, Pourkeshavarz M, Teimourian B, Ay MR, et al. Standard SPECT myocardial perfusion estimation from half-time acquisitions using deep convolutional residual neural networks. Journal of nuclear cardiology : official publication of the American Society of Nuclear Cardiology. 2020 Apr 28.

[28] Arabi H, Zeng G, Zheng G, Zaidi H. Novel adversarial semantic structure deep learning for MRI-guided attenuation correction in brain PET/MRI. European journal of nuclear medicine and molecular imaging. 2019;46:2746-59.